\documentclass[showpacs,preprintnumbers,amsmath,amssymb,APSl,prd,nofootinbib,superscriptaddress,12pt]{revtex4-2}
\usepackage{graphicx}
\usepackage{amssymb}
\usepackage{subcaption}
\usepackage{xcolor}
\usepackage{hyperref}
\usepackage{amsfonts}
\usepackage{bm}
\usepackage[a4paper, margin=1.6cm]{geometry}
\usepackage{mathrsfs}
\usepackage{soul}
\usepackage{makecell,multirow}
\usepackage{rotating}
\usepackage[utf8]{inputenc}
\usepackage{amsmath}
\usepackage{makecell}

\usepackage{amssymb}
\usepackage{tensor}
\usepackage{graphicx}
\usepackage{bigints}
\usepackage{bbm}
\usepackage{graphicx}
\usepackage{subcaption}
\usepackage{epsf}
\usepackage{bm}
\usepackage{amsmath}
\usepackage{amsfonts}
\usepackage{amssymb}
\usepackage{graphicx}
\usepackage{tabularx}
\usepackage{multirow}
\usepackage{color}%
\providecommand{\U}[1]{\protect\rule{.1in}{.1in}}

\newcommand{\ie}{\begin{equation}}
\newcommand{\fe}{\end{equation}}

\newcommand{\mincir}{\raise
-3.truept\hbox{\rlap{\hbox{$\sim$}}\raise4.truept\hbox{$<$}\ }}
\newcommand{\magcir}{\raise
-3.truept\hbox{\rlap{\hbox{$\sim$}}\raise4.truept\hbox{$>$}\ }}

\providecommand{\U}[1]{\protect\rule{.1in}{.1in}}

\usepackage{tikz,xcolor,hyperref}

\usepackage{hyperref}             
\hypersetup{
    colorlinks=true,              
    breaklinks=true,              
    citecolor=blue,               
    linkcolor=[rgb]{0,0.5,0.9},   
    urlcolor=red,                 
    filecolor=green               
}

\definecolor{lime}{HTML}{A6CE39}
\DeclareRobustCommand{\orcidicon}{%
	\begin{tikzpicture}
	\draw[lime, fill=lime] (0,0) 
	circle [radius=0.16] 
	node[white] {{\fontfamily{qag}\selectfont \tiny ID}};
	\draw[white, fill=white] (-0.0625,0.095) 
	circle [radius=0.007];
	\end{tikzpicture}
	\hspace{-2mm}
}

\foreach \x in {A, ..., Z}{%
	\expandafter\xdef\csname orcid\x\endcsname{\noexpand\href{https://orcid.org/\csname orcidauthor\x\endcsname}{\noexpand\orcidicon}}
}

\begin{document}

\title{\Large{Exploring phantom Dirac-Born-Infeld regular black hole via particle emission, wave scattering and geodesics}}

\author{N.~Heidari\orcidA{}}
\email{heidari.n@gmail.com (The corresponding author)}

\affiliation{Departamento de Física, Universidade Federal de Campina Grande Caixa Postal 10071, 58429-900 Campina Grande, Paraíba, Brazil.}

\affiliation{Center for Theoretical Physics, Khazar University, 41 Mehseti Street, Baku, AZ-1096, Azerbaijan.}

\affiliation{School of Physics, Damghan University, Damghan, 3671641167, Iran.}


\author{A.~A.~Ara\'{u}jo~Filho\orcidB{}}
\email{dilto@fisica.ufc.br}

\affiliation{Departamento de Física, Universidade Federal de Campina Grande Caixa Postal 10071, 58429-900 Campina Grande, Paraíba, Brazil.}
\affiliation{Center for Theoretical Physics, Khazar University, 41 Mehseti Street, Baku, AZ-1096, Azerbaijan.}
\affiliation{Departamento de Física, Universidade Federal da Paraíba, Caixa Postal 5008, 58051--970, João Pessoa, Paraíba,  Brazil.}

\author{Iarley P. Lobo\orcidD{}}
\email{lobofisica@gmail.com}

\affiliation{Departamento de Física, Universidade Federal da Paraíba, Caixa Postal 5008, 58051--970, João Pessoa, Paraíba,  Brazil.}
\affiliation{Departamento de Física, Universidade Federal de Campina Grande Caixa Postal 10071, 58429-900 Campina Grande, Paraíba, Brazil.}

\author{V. B. Bezerra\orcidG{}}
\email{valdir@fisica.ufpb.br}
\affiliation{Departamento de Física, Universidade Federal da Paraíba, Caixa Postal 5008, 58051--970, João Pessoa, Paraíba,  Brazil.}



\begin{abstract}

We examine particle creation, evaporation, scalar wave absorption and scattering, and geodesic motion in the asymptotically flat regular black hole supported by a phantom Dirac--Born--Infeld field. For massless bosonic and fermionic fields, the Bogoliubov transformations yield thermal spectra whose Hawking temperature decreases as the regular core becomes more prominent. Energy--conserving tunneling recovers the same temperature in the low--energy limit, whereas recoil and the DBI contribution introduce nonthermal corrections and suppress particle emission. In the high--frequency regime, the enlargement of the cross section does not compensate for the reduction in temperature, resulting in a lower luminosity and a longer evaporation time. A numerical partial wave analysis shows that the total scalar absorption increases with the regular core scale, approaches the horizon area at low frequencies, and oscillates around an enlarged geometric capture limit at high frequencies. The scattering phase shifts modify the multipolar amplitudes nonuniformly and displace the interference fringes toward larger angles. Furthermore, both null and timelike trajectories experience stronger deflection as the regular core contribution increases.

\end{abstract}

\maketitle

\pagebreak

\tableofcontents

\pagebreak

\section{Introduction{}}

Curvature singularities are among the clearest indications that the classical description of gravity has reached the boundary of its applicability \cite{penrose1965gravitational,penrose1978singularities,hawking1970singularities}. Regular black holes address this limitation by replacing the singular region with a smooth core while preserving a well--defined exterior and, within suitable ranges of the parameters, an event horizon \cite{Filho:2023qxu,Filho:2023voz}. Since Bardeen's original proposal \cite{Bardeen1968}, regular solutions have been obtained in general relativity coupled to nonlinear electrodynamics \cite{AyonBeatoGarcia1998,AyonBeatoGarcia1999,AyonBeatoGarcia2005,AraujoFilho:2026hun,Bronnikov2001,BalartVagenas2014,BurinskiiHildebrandt2002,Culetu2015Nonsingular,FanWang2016}, in braneworld and higher--curvature constructions, including $F(R)$ gravity \cite{shankaranarayanan2004non,HuLanMiao2023}, and through noncanonical scalar fields, vacuum-like sources, thin shells, fluids, dark matter distributions, or phenomenological mass profiles \cite{BronnikovFabris2006,Hayward2006,Dymnikova1992,uchikata2012new,VertogradovOvgun2025,KonoplyaZhidenko2026,heidari2026signatures,calza2025primordial}. A broader account of their theoretical foundations and phenomenology can be found in Ref.~\cite{bambi2023regular}.

Against this background, we consider the exact asymptotically flat solution obtained by Parvez and Shankaranarayanan in Einstein gravity minimally coupled to a Dirac--Born--Infeld (DBI) scalar field \cite{parvez2026exact}. The geometry is not introduced through an effective metric chosen independently of its source. Instead, the DBI action is combined with a regularity--motivated ansatz for the areal radius, while the field equations fix the lapse function and select the phantom branch of the scalar theory. The resulting spacetime is everywhere regular, carries nontrivial scalar hair, and approaches the Schwarzschild geometry asymptotically. Its departure from Schwarzschild is controlled by the dimensionless ratio $a/M$, where $a$ determines the size of the regular core.

The phantom character of the DBI field is essential to the construction because it relaxes the energy-condition hypotheses entering the classical singularity theorems. The global solution also contains a lower-mass boundary that was interpreted in Ref.~\cite{parvez2026exact} as a possible nonsingular evaporation remnant of order one gram. If sufficiently long--lived and produced with an adequate primordial abundance, such objects could contribute to dark matter. Their viability, however, depends not only on the absence of a curvature singularity but also on whether the radiative evolution actually approaches and terminates at this limiting configuration. This question is particularly delicate near $M\sim a$, where expansions based on $a/M\ll1$ are no longer controlled.

Particle creation provides a direct means of examining how the regular deformation modifies the horizon. The Hawking effect originates from the Bogoliubov mixing of positive-- and negative--frequency modes associated with distinct vacuum constructions in curved spacetime \cite{hawking1975particle,birrell1982quantum,wald1994quantum,AraujoFilho:2025rwr,AraujoFilho:2025hkm,AraujoFilho:2024ctw}. For a stationary nonextremal horizon, analytic continuation of the outgoing modes across the future horizon exposes the thermal factor and yields the corresponding Bose--Einstein and Fermi--Dirac distributions \cite{damour1976black,sannan1988heuristic}. The same near--horizon pole governs the complex path and tunneling formulations of the emission process \cite{srinivasan1999particle,parikh2000hawking,vanzo2011tunnelling}, with the Hamilton--Jacobi construction extending naturally to spin-$1/2$ fields \cite{kerner2008fermions}. On a fixed background, these methods isolate the temperature determined by the surface gravity. Once self gravitation and energy conservation are included, the horizon moves during emission and the spectrum acquires nonthermal corrections \cite{kraus1995self,parikh2000hawking}. The flux measured at infinity is further filtered by the spin--, frequency--, and angular momentum dependent greybody factors generated by the exterior curvature potential \cite{page1976emission}.

Studies devoted specifically to this spacetime remain recent. The original work examined its linear stability under massless scalar perturbations and discussed its possible evaporation endpoint \cite{parvez2026exact}. The Hawking temperature, entropy, heat capacity, sparsity, and spectral energy emission rate were subsequently considered in Ref.~\cite{ahmed2026hawking}. The quasinormal spectrum of a massive scalar field was obtained in Ref.~\cite{skvortsova2026massive}, whereas massless scalar, electromagnetic, and Dirac perturbations were analyzed in Ref.~\cite{lutfuouglu2026scalar}. Eikonal ringing, the photon sphere, the shadow, strong gravitational lensing, greybody estimates, and circular-orbit energetics were studied through geodesic optics in Ref.~\cite{abdullaev2026eikonal}. A calculation connecting bosonic and fermionic particle creation at the horizon to energy-conserving emission, evaporation, exact partial-wave cross sections, and particle motion has not yet been presented for this background.

In this work, we develop such a unified analysis. We initially derive the fixed-background Bogoliubov transformations for massless bosonic and fermionic fields, obtaining their Bose--Einstein and Fermi--Dirac occupation numbers and the common Hawking temperature determined by the outer horizon. We then impose energy conservation within the tunneling description by allowing the black hole mass to decrease along the classically forbidden trajectory. This procedure separates the low--energy thermal contribution from the recoil term and the additional nonthermal correction produced by the DBI scale. Having determined the mode occupations, we formulate the luminosity as a sum over particle species and angular channels, retain the greybody transmission probabilities, and derive the high--frequency corrections to the mass loss rate and evaporation time. The exterior wave response is obtained independently by solving the massless Klein--Gordon equation through a numerical partial-wave construction. From the resulting phase shifts and transmission coefficients, we calculate the partial and total absorption and scattering cross sections and examine their low- and high-frequency limits. Finally, we integrate the null and timelike geodesic equations to determine how the regular core affects light propagation and massive particle trajectories.

\section{Phantom DBI regular black hole geometry{\label{sec:setup}}}
\label{sec:geometry}

We consider the regular asymptotically flat solution obtained in Einstein gravity minimally coupled to a nonlinear Dirac--Born--Infeld scalar field. To fix our conventions, the underlying action can be written as \cite{parvez2026exact} 
\begin{equation} S=\int \mathrm{d}^{4}x\sqrt{-g}\left[ \frac{\mathcal{R}}{2\kappa^{2}} +\epsilon \Lambda^{4} \left( 1-\sqrt{1+\frac{2X}{\Lambda^{4}}} \right) \right], \label{eq:DBI_action} 
\end{equation} 
with 
\begin{equation} X=-\frac{1}{2}g^{\mu\nu}\nabla_{\mu}\phi\nabla_{\nu}\phi, \qquad \kappa^{2}=8\pi G. 
\end{equation}
Here $\Lambda$ is the measure of the non-linear DBI sector, and $\epsilon=\pm1$ is the sign indicating canonical and phantom branches. The phantom branch supports the regular solution, $\epsilon=+1$ \cite{parvez2026exact}, with the convention employed in Eq.~\eqref{eq:DBI_action}. 
The static and spherically symmetric line element is given by 

\begin{equation} 
\mathrm{d}s^{2} =-f(r)\mathrm{d}t^{2} +\frac{\mathrm{d}r^{2}}{f(r)} +R^{2}(r)\left(\mathrm{d}\theta^{2} +\sin^{2}\theta\,\mathrm{d}\varphi^{2}\right), \label{eq:DBImetric} 
\end{equation} 
where $r$ is a radial coordinate and $R(r)$ is the areal radius. 

The regularity condition is satisfied by 
\begin{equation} R(r)=\sqrt{r^{2}+a^{2}}. \label{eq:areal_radius} 
\end{equation} 

Thus $R(r)$ reaches the minimum value $R_{\min}=R(0)=a.$. The surface $r=0$ is then a two-sphere of finite area $4\pi a^{2}$ rather than a point where the areal radius becomes zero. The constant is of dimension of length and called the regular core scale.

For the choice Eq.~\eqref {eq:areal_radius}, a combination of the Einstein equations that does not involve the scalar profile becomes 
\begin{equation} \left(r^{2}+a^{2}\right)f''(r)-2f(r)+2=0. \label{eq:lapse} \end{equation} 

Requiring asymptotic flatness and identifying $M$ with the ADM mass fixes the lapse function uniquely to 

\begin{equation}
f(r) =1+\frac{3GM}{a} \left[ \frac{r}{a}-\left(1+\frac{r^{2}}{a^{2}}\right) \tan^{-1}\left(\frac{a}{r}\right) \right]. \label{eq:exactlapse} 
\end{equation}

From now on, we will work in geometrized units $G=c=\hbar=1$, unless specified otherwise. The metric functions behave asymptotically as \cite{parvez2026exact}
\begin{equation} \label{eq:R_asymptotic} R(r)=r+\frac{a^{2}}{2r} +\mathcal{O}\left(r^{-3}\right), \end{equation} \begin{equation} \label{eq:fasymptotic} f(r)= 1-\frac{2M}{r} +\frac{2Ma^{2}}{5r^{3}} +\mathcal{O}\left(r^{-5}\right). 
\end{equation} 

The ADM mass is therefore fixed by the usual $r^{-1}$ term, while the first correction to the Schwarzschild lapse appears at order $a^{2}/r^{3}$. It reduces in the limit $a\rightarrow0$ to the Schwarzschild geometry in a continuous way. In the regime $a/r\ll 1$ the lapse can be approximated as \begin{equation} f(r)\simeq 1-\frac{2M}{r} +\frac{2Ma^{2}}{5r^{3}}. \label{eq:approximate_lapse} \end{equation}
This modified form is a large--distance or small--core approximation, and its validity should not be assumed in the region $r\sim a$ without verification.


\section{Field excitations in quantum regimes}

Quantum emission from the phantom Dirac--Born--Infeld black hole will be examined at two complementary levels. The fixed-background calculation initially determines the mixing between positive-- and negative--frequency scalar modes and isolates the thermal factor generated by the event horizon. The tunneling description will subsequently enforce energy conservation by allowing the mass to change during emission.  


\subsection{Perturbations of a bosonic field}

\subsubsection{Thermal radiative emission}

We consider a massless scalar field on the fixed phantom Dirac--Born--Infeld background. In the regime $a/r\ll1$, the line element introduced previously reduces to
\ie
\nonumber
\label{eq:phantom-metric-particle-creation}
\mathrm{d}s^{2} \approx\, -f(r)\mathrm{d}t^{2} +\frac{\mathrm{d}r^{2}}{f(r)} +\mathcal{R}^{2}(r)\mathrm{d}\Omega^{2}, \qquad \mathcal{R}(r)=\sqrt{r^{2}+a^{2}},
\fe
with
\ie
\nonumber
\label{eq:phantom-lapse-expanded}
f(r) =1-\frac{2M}{r}+\frac{2Ma^{2}}{5r^{3}} +\mathcal{O}\!\left(\frac{a^{4}}{r^{5}}\right).
\fe
Here $M$ denotes the ADM mass, $a$ is the length scale of the regular core, $\mathcal{R}(r)$ is the areal radius, and $\mathrm{d}\Omega^{2}=\mathrm{d}\theta^{2} +\sin^{2}\theta\,\mathrm{d}\phi^{2}$ is the metric of the unit two--sphere. The approximation in Eq.~\eqref{eq:phantom-lapse-expanded} remains controlled near the outer horizon only when $a/M\ll1$. Although the lapse is expanded, the angular sector must retain $\mathcal{R}^{2}(r)=r^{2}+a^{2}$ because $r$ is not the areal radius of the phantom geometry

The scalar field operator $\Phi$ satisfies
\ie
\label{eq:phantom-kg}
\frac{1}{\sqrt{-g}}\, \partial_{\mu}\!\left( \sqrt{-g}\,g^{\mu\nu}\partial_{\nu}\Phi \right)=0,
\fe
where $g=\det(g_{\mu\nu})$. A complete quantization may be performed in two different mode bases~\cite{hawking1975particle,fulling1989aspects, hollands2015quantum,parker2009quantum,wald1994quantum}:
\ie
\label{eq:phantom-field-decomposition}
\Phi =\sum_i\left(f_i a_i+f_i^{*}a_i^{\dagger}\right) =\sum_i\left( p_i b_i+p_i^{*}b_i^{\dagger} +q_i c_i+q_i^{*}c_i^{\dagger} \right).
\fe
The modes $f_i$ have positive frequency on past null infinity. The set $p_i$ contains the outgoing modes that reach future null infinity, while $q_i$ completes the future basis with modes having no outgoing support there. The operators $a_i$, $b_i$, and $c_i$ annihilate the corresponding states, and the dagger denotes Hermitian conjugation.

The spherical harmonics $Y_{lm}(\theta,\phi)$ separate the angular dependence, with $l=0,1,\ldots$ and $m=-l,\ldots,l$. In the exterior region, the relevant modes can be expressed as
\ie
\label{eq:phantom-asymptotic-modes}
\begin{split}
f_{\omega'lm} &=\frac{\mathcal{F}_{\omega'l}(r)} {\sqrt{2\pi\omega'}\,\mathcal{R}(r)} e^{-i\omega'v}Y_{lm}(\theta,\phi),\\ p_{\omega lm} &=\frac{\mathcal{P}_{\omega l}(r)} {\sqrt{2\pi\omega}\,\mathcal{R}(r)} e^{-i\omega u}Y_{lm}(\theta,\phi).
\end{split}
\fe
The quantities $\omega'>0$ and $\omega>0$ are the Killing frequencies of the
ingoing and outgoing bases, respectively, while
$\mathcal{F}_{\omega'l}$ and $\mathcal{P}_{\omega l}$ denote their radial
profiles. The advanced and retarded coordinates are
\ie
\label{eq:phantom-null-coordinates}
v=t+r^{*}, \qquad u=t-r^{*}, \qquad \frac{\mathrm{d}r^{*}}{\mathrm{d}r}=\frac{1}{f(r)}.
\fe

The tortoise coordinate can be written without displaying the lengthy radical
form returned by a direct integration. Multiplying
Eq.~\eqref{eq:phantom-lapse-expanded} by $r^{3}$ defines the monic horizon
polynomial
\ie
\label{eq:phantom-horizon-polynomial}
P(r) \equiv r^{3}-2Mr^{2}+\frac{2Ma^{2}}{5} =(r-r_{+})(r-r_{-})(r-r_{\mathrm n}),
\fe
where the three algebraic roots are
\ie
\label{eq:phantom-three-roots}
\begin{split}
r_{+} &=\frac{2M}{3}+\frac{4M}{3}\cos\vartheta,\\ r_{-} &=\frac{2M}{3}+\frac{4M}{3} \cos\!\left(\vartheta-\frac{2\pi}{3}\right),\\ r_{\mathrm n} &=\frac{2M}{3}+\frac{4M}{3} \cos\!\left(\vartheta-\frac{4\pi}{3}\right), \end{split} \qquad \vartheta \equiv\frac{1}{3} \arccos\!\left(1-\frac{27a^{2}}{40M^{2}}\right).
\fe

For distinct roots, partial fractions give
\ie
\label{eq:phantom-tortoise-roots}
r^{*} =r+\sum_{i\in\{+,-,\mathrm n\}} \frac{r_i^{2}}{3r_i-4M}\ln|r-r_i|+C_{*},
\fe
where $C_{*}$ is an integration constant. The coefficient of each logarithm is
$1/f'(r_i)=r_i^{2}/(3r_i-4M)$. Expanding the outer root gives
\ie
\label{eq:phantom-horizon-series}
r_{\mathrm h} \equiv r_{+} =2M-\frac{a^{2}}{10M} +\mathcal{O}\!\left(\frac{a^{4}}{M^{3}}\right).
\fe
The surface gravity of this nonextremal Killing horizon is
\ie
\label{eq:phantom-kappa}
\kappa_{\mathrm h} \equiv\frac{f'(r_{\mathrm h})}{2} =\frac{3r_{\mathrm h}-4M}{2r_{\mathrm h}^{2}} =\frac{1}{4M} \left(1-\frac{a^{2}}{20M^{2}}\right) +\mathcal{O}\!\left(\frac{a^{4}}{M^{5}}\right).
\fe
It follows that
\ie
\label{eq:phantom-inverse-kappa}
\frac{1}{\kappa_{\mathrm h}} =4M+\frac{a^{2}}{5M} +\mathcal{O}\!\left(\frac{a^{4}}{M^{3}}\right),
\fe
and the only singular contribution to the tortoise coordinate is
\ie
\label{eq:phantom-tortoise-near-horizon}
r^{*} =\frac{1}{2\kappa_{\mathrm h}}\ln|r-r_{\mathrm h}| +\text{terms regular at }r=r_{\mathrm h}.
\fe

We next follow a future--directed radial null geodesic toward the horizon. Let $\lambda$ be an affine parameter and let $E\equiv-p_t>0$ denote its conserved Killing energy. The null condition and time translation symmetry give
\ie
\label{eq:phantom-null-geodesic-relations}
E=f(r)\frac{\mathrm{d}t}{\mathrm{d}\lambda}, \qquad \left(\frac{\mathrm{d}r}{\mathrm{d}\lambda}\right)^{2}=E^{2}.
\fe
The ingoing branch satisfies
$\mathrm{d}r/\mathrm{d}\lambda=-E$. Setting $\lambda=0$ at the outer horizon
leads to
\ie
\label{eq:phantom-null-trajectory}
r(\lambda)=r_{\mathrm h}-E\lambda.
\fe
The exterior segment corresponds to $\lambda<0$. Along this consideration, $v$ remains constant, whereas the retarded coordinate obeys
\ie
\label{eq:phantom-u-affine}
\frac{\mathrm{d}u}{\mathrm{d}\lambda}
=\frac{2E}{f(r(\lambda))}.
\fe
Using Eqs.~\eqref{eq:phantom-lapse-expanded} and \eqref{eq:phantom-null-trajectory}, the integration through order $a^{2}$ yields
\ie
\label{eq:phantom-u-full}
\begin{split}
u(\lambda,a) =u_{0} &+2E\lambda -\left(4M+\frac{a^{2}}{5M}\right) \ln\!\left(-\frac{\lambda}{C_{\lambda}}\right)\\ &+\frac{a^{2}}{5M} \ln\!\left(\frac{2M-E\lambda}{2M}\right) +\mathcal{O}\!\left(\frac{a^{4}}{M^{3}}\right).
\end{split}
\fe
The constant $u_{0}$ fixes the origin of the retarded coordinate, and $C_{\lambda}>0$ has the same dimension as $\lambda$. The second logarithm in Eq.~\eqref{eq:phantom-u-full} is finite at $\lambda=0$ and contributes only to the part analytic at the horizon. The singular term becomes
\ie
\label{eq:phantom-u-near-horizon}
u(\lambda,a) =u_{0} -\left(4M+\frac{a^{2}}{5M}\right) \ln\!\left(-\frac{\lambda}{C_{\lambda}}\right) +\mathcal{O}(\lambda) +\mathcal{O}\!\left(\frac{a^{4}}{M^{3}}\right).
\fe

For the family of rays that narrowly escapes horizon formation, geometric optics relates the positive exterior separation $\ell\equiv-\lambda$ to the advanced coordinate according to~\cite{calmet2023quantum}
\ie
\ell=\frac{v_{0}-v}{D}, \qquad   D>0.
\fe
Here $v_{0}$ labels the last ingoing ray that can still emerge before the horizon forms, while $D$ fixes the normalization of the affine parameter. Defining $C_v\equiv D C_{\lambda}>0$, we have
\ie
\label{eq:phantom-ray-tracing-explicit}
u(v) =u_{0} -\left(4M+\frac{a^{2}}{5M}\right) \ln\!\left(\frac{v_{0}-v}{C_v}\right) +\mathcal{O}(v_{0}-v) +\mathcal{O}\!\left(\frac{a^{4}}{M^{3}}\right), \qquad v<v_{0}.
\fe
The combination
\ie
\label{eq:phantom-Xi-definition}
\Xi(M,a) \equiv4M+\frac{a^{2}}{5M}
\fe
will be used only as a shorthand. Within the retained approximation, $\Xi=1/\kappa_{\mathrm h}$. An outgoing mode traced back to past null infinity then has the form
\ie
\label{eq:phantom-traced-mode}
p_{\omega}(v) =K_{\omega}\, \Theta(v_{0}-v) \left(\frac{v_{0}-v}{C_v}\right)^{i\omega\Xi(M,a)}.
\fe
The Heaviside function $\Theta(v_{0}-v)$ restricts the mode to rays with $v<v_{0}$. The constant $K_{\omega}$ contains the mode normalization, the regular radial contribution, and the phase $e^{-i\omega u_{0}}$.

The outgoing mode is expanded in the ingoing basis as
\ie
\label{eq:phantom-bogoliubov-expansion}
p_{\omega} =\int_{0}^{\infty} \left[ \alpha_{\omega\omega'}f_{\omega'} +\beta_{\omega\omega'}f_{\omega'}^{*} \right]\mathrm{d}\omega'.
\fe
Let $y\equiv v_{0}-v>0$. With the phase convention adopted in Eq.~\eqref{eq:phantom-asymptotic-modes}, the coefficients follow from
\ie
\label{eq:phantom-bogoliubov-integrals}
\begin{split}
\alpha_{\omega\omega'} &=\frac{K_{\omega}}{2\pi} \sqrt{\frac{\omega'}{\omega}}\, e^{i\omega'v_{0}} \lim_{\varepsilon\to0^{+}} \int_{0}^{\infty}\mathrm{d}y\, e^{-(\varepsilon+i\omega')y} \left(\frac{y}{C_v}\right)^{i\omega\Xi(M,a)},\\ \beta_{\omega\omega'} &=-\frac{K_{\omega}}{2\pi} \sqrt{\frac{\omega'}{\omega}}\, e^{-i\omega'v_{0}} \lim_{\varepsilon\to0^{+}} \int_{0}^{\infty}\mathrm{d}y\, e^{-(\varepsilon-i\omega')y} \left(\frac{y}{C_v}\right)^{i\omega\Xi(M,a)}.
\end{split}
\fe
The positive regulator $\varepsilon$ specifies the convergence prescription. Using
\ie
\int_{0}^{\infty} y^{s-1}e^{-zy}\mathrm{d}y =z^{-s}\Gamma_{\mathrm E}(s), \qquad \operatorname{Re}(z)>0,
\fe
where $\Gamma_{\mathrm E}$ denotes the Euler Gamma function, the coefficients become
\ie
\label{eq:phantom-alpha-explicit}
\begin{aligned}
\alpha_{\omega\omega'}(M,a) =&-\frac{iK_{\omega}}{2\pi\sqrt{\omega\omega'}}\, e^{i\omega'v_{0}}\, \exp\!\left[ \frac{\pi\omega}{2} \left(4M+\frac{a^{2}}{5M}\right) \right]\\ &\times \Gamma_{\mathrm E}\!\left[ 1+i\omega\left(4M+\frac{a^{2}}{5M}\right) \right] \left(C_v\omega'\right)^{ -i\omega\left(4M+\frac{a^{2}}{5M}\right)}
\end{aligned}
\fe
and
\ie
\label{eq:phantom-beta-explicit}
\begin{aligned}
\beta_{\omega\omega'}(M,a) =&-\frac{iK_{\omega}}{2\pi\sqrt{\omega\omega'}}\, e^{-i\omega'v_{0}}\, \exp\!\left[ -\frac{\pi\omega}{2} \left(4M+\frac{a^{2}}{5M}\right) \right]\\ &\times \Gamma_{\mathrm E}\!\left[ 1+i\omega\left(4M+\frac{a^{2}}{5M}\right) \right] \left(C_v\omega'\right)^{ -i\omega\left(4M+\frac{a^{2}}{5M}\right)} .
\end{aligned}
\fe
It is worth mentioning that their phases depend on the convention chosen for the mode
basis, but their absolute values and ratio do not. Moreover, the identity
\ie
\left|\Gamma_{\mathrm E}(1+ix)\right|^{2} =\frac{\pi x}{\sinh(\pi x)}
\fe
gives the explicit squared amplitudes
\ie
\label{eq:phantom-alpha-absolute}
\left|\alpha_{\omega\omega'}\right|^{2} =\frac{|K_{\omega}|^{2}}{2\pi\omega'} \left(4M+\frac{a^{2}}{5M}\right) \frac{1}{ 1-\exp\!\left[ -2\pi\omega\left(4M+\frac{a^{2}}{5M}\right) \right]},
\fe
and
\ie
\label{eq:phantom-beta-absolute}
\left|\beta_{\omega\omega'}\right|^{2} =\frac{|K_{\omega}|^{2}}{2\pi\omega'} \left(4M+\frac{a^{2}}{5M}\right) \frac{1}{ \exp\!\left[ 2\pi\omega\left(4M+\frac{a^{2}}{5M}\right) \right]-1}.
\fe
The normalization--dependent prefactor cancels from the quotient, leaving
\ie
\label{eq:phantom-bogoliubov-ratio-explicit}
\frac{ \left|\alpha_{\omega\omega'}\right|^{2} }{ \left|\beta_{\omega\omega'}\right|^{2} } =\exp\!\left[ \left(8\pi M+\frac{2\pi a^{2}}{5M}\right)\omega \right].
\fe
The mass and the DBI scale therefore enter the mode mixing through the single near-horizon combination $\Xi(M,a)$. The separate coefficients also contain $v_{0}$, $C_v$, and $K_{\omega}$, which encode the collapse history, the affine normalization, and the normalization of the mode. These quantities do not modify the thermal quotient.

The number of quanta created in $[\omega,\omega+\mathrm{d}\omega]$ follows from Eq.~\eqref{eq:phantom-bogoliubov-ratio-explicit}:
\ie
\label{eq:phantom-planck-factor-explicit}
\mathrm{d}\mathcal{N}^{(\mathrm H)}_{\omega} =\frac{\mathrm{d}\omega}{2\pi} \frac{1}{ \exp\!\left[ \left(8\pi M+\frac{2\pi a^{2}}{5M}\right)\omega \right]-1}.
\fe
Comparison with the Planck factor determines
\ie
\label{eq:phantom-temperature-explicit}
T_{\mathrm H} =\frac{\kappa_{\mathrm h}}{2\pi} =\frac{1}{2\pi\Xi(M,a)} =\frac{1}{8\pi M} \left(1-\frac{a^{2}}{20M^{2}}\right) +\mathcal{O}\!\left(\frac{a^{4}}{M^{5}}\right).
\fe
At fixed ADM mass, a nonzero $a$ lowers the temperature and suppresses the occupation of every mode with $\omega>0$. This conclusion refers to particle creation on the fixed background. The number flux measured at infinity also contains the transmission probability $\mathscr{T}_{\omega l}\in[0,1]$ for each partial wave:
\ie
\label{eq:phantom-greybody-flux}
\frac{\mathrm{d}^{2}\mathcal{N}_{\infty}} {\mathrm{d}t\,\mathrm{d}\omega} =\frac{1}{2\pi} \sum_{l=0}^{\infty}(2l+1) \frac{\mathscr{T}_{\omega l}}{ \exp(\omega/T_{\mathrm H})-1}.
\fe
The factor $\mathscr{T}_{\omega l}$ produces the greybody deformation through scattering outside the horizon. It must not be confused with the Euler Gamma function in Eqs.~\eqref{eq:phantom-alpha-explicit} and \eqref{eq:phantom-beta-explicit}, and it does not alter the Planckian near--horizon ratio. Nonthermal corrections associated with energy conservation arise only after the mass is allowed to vary in the tunneling calculation.

Finally, the perturbative result Eq.~\eqref{eq:phantom-temperature-explicit} cannot be used to determine the endpoint of evaporation. Setting it formally to zero gives $a^{2}=20M^{2}$, in direct conflict with the assumption $a/M\ll1$. The extremal relic of the complete phantom DBI geometry, $M_{\mathrm{relic}}=2a/(3\pi)$ in the present units, follows from the unexpanded lapse and cannot be recovered by extrapolating Eq.~\eqref{eq:phantom-lapse-expanded}.


\section{Field excitations in quantum regimes}

Quantum emission from the phantom Dirac--Born--Infeld black hole will be examined at two complementary levels. The fixed background calculation initially determines the mixing between positive and negative frequency scalar modes and isolates the thermal factor generated by the event horizon. The tunneling description will subsequently enforce energy conservation by allowing the mass to change during emission. The fermionic sector will be treated through the near-horizon Hamilton--Jacobi equation, which is sufficient to determine its leading occupation number. These three calculations probe different aspects of the same geometry and must be distinguished when interpreting thermality, backreaction, and greybody effects.


\subsection{Perturbations of a bosonic field}


\subsubsection{Thermal radiative emission}

We consider a massless scalar field on the fixed phantom Dirac--Born--Infeld background. In the regime $a/r\ll1$, the line element introduced previously reduces to
\ie
\label{eq:phantom-metric-particle-creation}
\mathrm{d}s^{2} =-f(r)\mathrm{d}t^{2} +\frac{\mathrm{d}r^{2}}{f(r)} +\mathcal{R}^{2}(r)\mathrm{d}\Omega^{2}, \qquad \mathcal{R}(r)=\sqrt{r^{2}+a^{2}},
\fe
with
\ie
\label{eq:phantom-lapse-expanded}
f(r) =1-\frac{2M}{r}+\frac{2Ma^{2}}{5r^{3}} +\mathcal{O}\!\left(\frac{a^{4}}{r^{5}}\right).
\fe
Here $M$ denotes the ADM mass, $a$ is the length scale of the regular core, $\mathcal{R}(r)$ is the areal radius, and $\mathrm{d}\Omega^{2}=\mathrm{d}\theta^{2} +\sin^{2}\theta\,\mathrm{d}\phi^{2}$ is the metric of the unit two--sphere. The approximation in Eq.~\eqref{eq:phantom-lapse-expanded} remains controlled near the outer horizon only when $a/M\ll1$. Although the lapse is expanded, the angular sector must retain $\mathcal{R}^{2}(r)=r^{2}+a^{2}$ because $r$ is not the areal radius of the phantom geometry.

The scalar field operator $\Phi$ satisfies
\ie
\label{eq:phantom-kg}
\frac{1}{\sqrt{-g}}\, \partial_{\mu}\!\left( \sqrt{-g}\,g^{\mu\nu}\partial_{\nu}\Phi \right)=0,
\fe
where $g=\det(g_{\mu\nu})$. A complete quantization may be performed in two different mode bases~\cite{hawking1975particle,fulling1989aspects, hollands2015quantum,parker2009quantum,wald1994quantum}:
\ie
\label{eq:phantom-field-decomposition}
\Phi =\sum_i\left(f_i a_i+f_i^{*}a_i^{\dagger}\right) =\sum_i\left( p_i b_i+p_i^{*}b_i^{\dagger} +q_i c_i+q_i^{*}c_i^{\dagger} \right).
\fe
The modes $f_i$ have positive frequency on past null infinity. The set $p_i$ contains the outgoing modes that reach future null infinity, while $q_i$ completes the future basis with modes having no outgoing support there. The operators $a_i$, $b_i$, and $c_i$ annihilate the corresponding states, and the dagger denotes Hermitian conjugation.

The spherical harmonics $Y_{lm}(\theta,\phi)$ separate the angular dependence, with $l=0,1,\ldots$ and $m=-l,\ldots,l$. In the exterior region, the relevant modes can be expressed as
\ie
\label{eq:phantom-asymptotic-modes}
\begin{split}
f_{\omega'lm} &=\frac{\mathcal{F}_{\omega'l}(r)} {\sqrt{2\pi\omega'}\,\mathcal{R}(r)} e^{-i\omega'v}Y_{lm}(\theta,\phi),\\ p_{\omega lm} &=\frac{\mathcal{P}_{\omega l}(r)} {\sqrt{2\pi\omega}\,\mathcal{R}(r)} e^{-i\omega u}Y_{lm}(\theta,\phi).
\end{split}
\fe
The quantities $\omega'>0$ and $\omega>0$ are the Killing frequencies of the ingoing and outgoing bases, respectively, while $\mathcal{F}_{\omega'l}$ and $\mathcal{P}_{\omega l}$ denote their radial profiles. The advanced and retarded coordinates are
\ie
\label{eq:phantom-null-coordinates}
v=t+r^{*}, \qquad u=t-r^{*}, \qquad \frac{\mathrm{d}r^{*}}{\mathrm{d}r}=\frac{1}{f(r)}.
\fe

The tortoise coordinate can be written without displaying the lengthy radical form returned by a direct integration. Multiplying Eq.~\eqref{eq:phantom-lapse-expanded} by $r^{3}$ defines the monic horizon polynomial
\ie
\label{eq:phantom-horizon-polynomial}
P(r) \equiv r^{3}-2Mr^{2}+\frac{2Ma^{2}}{5} =(r-r_{+})(r-r_{-})(r-r_{\mathrm n}),
\fe
where the three algebraic roots are
\ie
\label{eq:phantom-three-roots}
\begin{split}
r_{+} &=\frac{2M}{3}+\frac{4M}{3}\cos\vartheta,\\ r_{-} &=\frac{2M}{3}+\frac{4M}{3} \cos\!\left(\vartheta-\frac{2\pi}{3}\right),\\ r_{\mathrm n} &=\frac{2M}{3}+\frac{4M}{3} \cos\!\left(\vartheta-\frac{4\pi}{3}\right), \end{split} \qquad \vartheta \equiv\frac{1}{3} \arccos\!\left(1-\frac{27a^{2}}{40M^{2}}\right).
\fe

Moreover, for distinct roots, partial fractions give
\ie
\label{eq:phantom-tortoise-roots}
r^{*} =r+\sum_{i\in\{+,-,\mathrm n\}} \frac{r_i^{2}}{3r_i-4M}\ln|r-r_i|+C_{*},
\fe
where $C_{*}$ is an integration constant. The coefficient of each logarithm is $1/f'(r_i)=r_i^{2}/(3r_i-4M)$. Expanding the outer root gives
\ie
\label{eq:phantom-horizon-series}
r_{\mathrm h} \equiv r_{+} =2M-\frac{a^{2}}{10M} +\mathcal{O}\!\left(\frac{a^{4}}{M^{3}}\right).
\fe
The surface gravity of this nonextremal Killing horizon is
\ie
\label{eq:phantom-kappa}
\kappa_{\mathrm h} \equiv\frac{f'(r_{\mathrm h})}{2} =\frac{3r_{\mathrm h}-4M}{2r_{\mathrm h}^{2}} =\frac{1}{4M} \left(1-\frac{a^{2}}{20M^{2}}\right) +\mathcal{O}\!\left(\frac{a^{4}}{M^{5}}\right).
\fe
It follows that
\ie
\label{eq:phantom-inverse-kappa}
\frac{1}{\kappa_{\mathrm h}} =4M+\frac{a^{2}}{5M} +\mathcal{O}\!\left(\frac{a^{4}}{M^{3}}\right),
\fe
and the only singular contribution to the tortoise coordinate is
\ie
\label{eq:phantom-tortoise-near-horizon}
r^{*} =\frac{1}{2\kappa_{\mathrm h}}\ln|r-r_{\mathrm h}| +\text{terms regular at }r=r_{\mathrm h}.
\fe

We next follow a future directed radial null geodesic toward the horizon. Let $\lambda$ be an affine parameter and let $E\equiv-p_t>0$ denote its conserved Killing energy. The null condition and time translation symmetry give
\ie
\label{eq:phantom-null-geodesic-relations}
E=f(r)\frac{\mathrm{d}t}{\mathrm{d}\lambda}, \qquad \left(\frac{\mathrm{d}r}{\mathrm{d}\lambda}\right)^{2} = E^{2}.
\fe
The ingoing branch satisfies $\mathrm{d}r/\mathrm{d}\lambda=-E$. Setting $\lambda=0$ at the outer horizon leads to
\ie
\label{eq:phantom-null-trajectory}
r(\lambda)=r_{\mathrm h}-E\lambda.
\fe
The exterior segment corresponds to $\lambda<0$. Along this, $v$ remains constant, whereas the retarded coordinate obeys
\ie
\label{eq:phantom-u-affine}
\frac{\mathrm{d}u}{\mathrm{d}\lambda} =\frac{2E}{f(r(\lambda))}.
\fe
Using Eqs.~\eqref{eq:phantom-lapse-expanded} and \eqref{eq:phantom-null-trajectory}, the integration through order $a^{2}$ yields
\ie
\label{eq:phantom-u-full}
\begin{split}
u(\lambda,a) =u_{0} &+2E\lambda -\left(4M+\frac{a^{2}}{5M}\right) \ln\!\left(-\frac{\lambda}{C_{\lambda}}\right)\\ &+\frac{a^{2}}{5M} \ln\!\left(\frac{2M-E\lambda}{2M}\right) +\mathcal{O}\!\left(\frac{a^{4}}{M^{3}}\right).
\end{split}
\fe
The constant $u_{0}$ fixes the origin of the retarded coordinate, and $C_{\lambda}>0$ has the same dimension as $\lambda$. The second logarithm in Eq.~\eqref{eq:phantom-u-full} is finite at $\lambda=0$ and contributes only to the part analytic at the horizon. The singular term becomes
\ie
\label{eq:phantom-u-near-horizon}
u(\lambda,a) =u_{0} -\left(4M+\frac{a^{2}}{5M}\right) \ln\!\left(-\frac{\lambda}{C_{\lambda}}\right) +\mathcal{O}(\lambda) +\mathcal{O}\!\left(\frac{a^{4}}{M^{3}}\right).
\fe

For the family of rays that narrowly escapes horizon formation, geometric optics relates the positive exterior separation $\ell\equiv-\lambda$ to the advanced coordinate according to~\cite{calmet2023quantum}
\ie
\ell=\frac{v_{0}-v}{D}, \qquad D>0.
\fe
Here $v_{0}$ labels the last ingoing ray that can still emerge before the horizon forms, while $D$ fixes the normalization of the affine parameter. Defining $C_v\equiv D C_{\lambda}>0$, we obtain
\ie
\label{eq:phantom-ray-tracing-explicit}
u(v) = u_{0} -\left(4M+\frac{a^{2}}{5M}\right) \ln\!\left(\frac{v_{0}-v}{C_v}\right) +\mathcal{O}(v_{0}-v) +\mathcal{O}\!\left(\frac{a^{4}}{M^{3}}\right), \qquad v<v_{0}.
\fe
The combination
\ie
\label{eq:phantom-Xi-definition}
\Xi(M,a) \equiv4M+\frac{a^{2}}{5M}
\fe
will be used only as a shorthand. Within the retained approximation, $\Xi=1/\kappa_{\mathrm h}$. An outgoing mode traced back to past null infinity then has the form
\ie
\label{eq:phantom-traced-mode}
p_{\omega}(v) =K_{\omega}\, \Theta(v_{0}-v) \left(\frac{v_{0}-v}{C_v}\right)^{i\omega\Xi(M,a)}.
\fe
The Heaviside function $\Theta(v_{0}-v)$ restricts the mode to rays with $v<v_{0}$, as we should expect. The constant $K_{\omega}$ contains the mode normalization, the regular radial contribution, and the phase $e^{-i\omega u_{0}}$.

The outgoing mode is expanded in the ingoing basis as
\ie
\label{eq:phantom-bogoliubov-expansion}
p_{\omega} =\int_{0}^{\infty} \left[ \alpha_{\omega\omega'}f_{\omega'} +\beta_{\omega\omega'}f_{\omega'}^{*} \right]\mathrm{d}\omega'.
\fe
Let $y\equiv v_{0}-v>0$. With the phase convention adopted in Eq.~\eqref{eq:phantom-asymptotic-modes}, the coefficients follow from
\ie
\label{eq:phantom-bogoliubov-integrals}
\begin{split}
\alpha_{\omega\omega'} &=\frac{K_{\omega}}{2\pi} \sqrt{\frac{\omega'}{\omega}}\, e^{i\omega'v_{0}} \lim_{\varepsilon\to0^{+}} \int_{0}^{\infty}\mathrm{d}y\, e^{-(\varepsilon+i\omega')y} \left(\frac{y}{C_v}\right)^{i\omega\Xi(M,a)},\\ \beta_{\omega\omega'} &=-\frac{K_{\omega}}{2\pi} \sqrt{\frac{\omega'}{\omega}}\, e^{-i\omega'v_{0}} \lim_{\varepsilon\to0^{+}} \int_{0}^{\infty}\mathrm{d}y\, e^{-(\varepsilon-i\omega')y} \left(\frac{y}{C_v}\right)^{i\omega\Xi(M,a)}.
\end{split}
\fe
The positive regulator $\varepsilon$ specifies the convergence prescription. Using
\ie
\int_{0}^{\infty} y^{s-1}e^{-zy}\mathrm{d}y =z^{-s}\Gamma_{\mathrm E}(s), \qquad \operatorname{Re}(z)>0,
\fe
where $\Gamma_{\mathrm E}$ denotes the Euler Gamma function, the coefficients become
\ie
\label{eq:phantom-alpha-explicit}
\begin{aligned}
\alpha_{\omega\omega'}(M,a) = & -\frac{iK_{\omega}}{2\pi\sqrt{\omega\omega'}}\, e^{i\omega'v_{0}}\, \exp\!\left[ \frac{\pi\omega}{2} \left(4M+\frac{a^{2}}{5M}\right) \right]\\ &\times \Gamma_{\mathrm E}\!\left[ 1+i\omega\left(4M+\frac{a^{2}}{5M}\right) \right] \left(C_v\omega'\right)^{ -i\omega\left(4M+\frac{a^{2}}{5M}\right)}
\end{aligned}
\fe
and
\ie
\label{eq:phantom-beta-explicit}
\begin{aligned}
\beta_{\omega\omega'}(M,a) =&-\frac{iK_{\omega}}{2\pi\sqrt{\omega\omega'}}\, e^{-i\omega'v_{0}}\, \exp\!\left[ -\frac{\pi\omega}{2} \left(4M+\frac{a^{2}}{5M}\right) \right]\\ &\times \Gamma_{\mathrm E}\!\left[ 1+i\omega\left(4M+\frac{a^{2}}{5M}\right) \right] \left(C_v\omega'\right)^{ -i\omega\left(4M+\frac{a^{2}}{5M}\right)}.
\end{aligned}
\fe
Eq.~\eqref{eq:phantom-alpha-explicit} and \eqref{eq:phantom-beta-explicit} are understood consistently through $\mathcal{O}(a^{2})$. Their phases depend on the convention chosen for the mode basis, but their absolute values and ratio do not.

The identity
\ie
\left|\Gamma_{\mathrm E}(1+ix)\right|^{2} =\frac{\pi x}{\sinh(\pi x)}
\fe
gives the explicit squared amplitudes
\ie
\label{eq:phantom-alpha-absolute}
\left|\alpha_{\omega\omega'}\right|^{2} =\frac{|K_{\omega}|^{2}}{2\pi\omega'} \left(4M+\frac{a^{2}}{5M}\right) \frac{1}{ 1-\exp\!\left[ -2\pi\omega\left(4M+\frac{a^{2}}{5M}\right) \right]},
\fe
and
\ie
\label{eq:phantom-beta-absolute}
\left|\beta_{\omega\omega'}\right|^{2} =\frac{|K_{\omega}|^{2}}{2\pi\omega'} \left(4M+\frac{a^{2}}{5M}\right) \frac{1}{ \exp\!\left[ 2\pi\omega\left(4M+\frac{a^{2}}{5M}\right) \right]-1}.
\fe
The normalization--dependent prefactor cancels from the quotient, leaving
\ie
\label{eq:phantom-bogoliubov-ratio-explicit}
\frac{ \left|\alpha_{\omega\omega'}\right|^{2} }{ \left|\beta_{\omega\omega'}\right|^{2} } =\exp\!\left[ \left(8\pi M+\frac{2\pi a^{2}}{5M}\right)\omega \right].
\fe
The mass and the DBI scale therefore enter the mode mixing through the single near horizon combination $\Xi(M,a)$. The separate coefficients also contain $v_{0}$, $C_v$, and $K_{\omega}$, which encode the collapse history, the affine normalization, and the normalization of the mode. These quantities do not modify the thermal quotient.

The number of quanta created in $[\omega,\omega+\mathrm{d}\omega]$ follows from Eq.~\eqref{eq:phantom-bogoliubov-ratio-explicit}:
\ie
\label{eq:phantom-planck-factor-explicit}
\mathrm{d}\mathcal{N}^{(\mathrm H)}_{\omega} =\frac{\mathrm{d}\omega}{2\pi} \frac{1}{ \exp\!\left[ \left(8\pi M+\frac{2\pi a^{2}}{5M}\right)\omega \right]-1}.
\fe
Comparison with the Planck factor determines
\ie
\label{eq:phantom-temperature-explicit}
T_{\mathrm H} =\frac{\kappa_{\mathrm h}}{2\pi} =\frac{1}{2\pi\Xi(M,a)} =\frac{1}{8\pi M} \left(1-\frac{a^{2}}{20M^{2}}\right) +\mathcal{O}\!\left(\frac{a^{4}}{M^{5}}\right).
\fe
At fixed ADM mass, a nonzero $a$ lowers the temperature and suppresses the occupation of every mode with $\omega>0$. This conclusion refers to particle creation on the fixed background. The number flux measured at infinity also contains the transmission probability $\mathscr{T}_{\omega l}\in[0,1]$ for each partial wave:
\ie
\label{eq:phantom-greybody-flux}
\frac{\mathrm{d}^{2}\mathcal{N}_{\infty}} {\mathrm{d}t\,\mathrm{d}\omega} =\frac{1}{2\pi} \sum_{l=0}^{\infty}(2l+1) \frac{\mathscr{T}_{\omega l}}{ \exp(\omega/T_{\mathrm H})-1}.
\fe
The factor $\mathscr{T}_{\omega l}$ produces the greybody deformation through scattering outside the horizon. It must not be confused with the Euler Gamma function in Eqs.~\eqref{eq:phantom-alpha-explicit} and \eqref{eq:phantom-beta-explicit}, and it does not alter the Planckian near horizon ratio. Nonthermal corrections associated with energy conservation arise only after the mass is allowed to vary in the tunneling calculation.

Finally, the perturbative result Eq.~\eqref{eq:phantom-temperature-explicit} cannot be used to determine the endpoint of evaporation. Setting it formally to zero gives $a^{2}=20M^{2}$, in direct conflict with the assumption $a/M\ll1$. The extremal relic of the complete phantom DBI geometry, $M_{\mathrm{relic}}=2a/(3\pi)$ in the present units, follows from the unexpanded lapse and cannot be recovered by extrapolating Eq.~\eqref{eq:phantom-lapse-expanded}.


\subsubsection{Tunneling-based analysis}

The fixed--background calculation determines the thermal factor generated by the outer horizon, but it does not account for the energy removed from the black hole. We now describe the emitted quantum as a self--gravitating, outgoing shell and impose energy conservation during the entire trajectory \cite{parikh2004energy,vanzo2011tunnelling,calmet2023quantum}. We work in units $G=c=\hbar=k_{\mathrm B}=1$ and consider neutral emission, so the DBI scale $a$ remains fixed. If the shell has acquired an energy $\omega'$, the instantaneous mass entering the geometry is
\ie
\label{eq:phantom-tunnel-running-mass}
m=M-\omega', \qquad 0\leq\omega'\leq\omega<M.
\fe
The perturbative geometry is reliable along the complete tunneling path only when $a/(M-\omega)\ll1$.

It is convenient to replace the static time $t_{\mathrm s}$ by a Painlev\'e--Gullstrand time $\tau$. At the accuracy retained here,
\ie
\nonumber
\label{eq:phantom-tunnel-static-metric}
\mathrm{d}s^{2} =-f(r;m)\mathrm{d}t_{\mathrm s}^{2} +\frac{\mathrm{d}r^{2}}{f(r;m)} +\mathcal{R}^{2}(r)\mathrm{d}\Omega^{2}, \qquad f(r;m)=1-\frac{2m}{r}+\frac{2ma^{2}}{5r^{3}}.
\fe
In this manner, the transformation
\ie
\label{eq:phantom-tunnel-pg-transformation}
\mathrm{d}t_{\mathrm s} =\mathrm{d}\tau -\frac{\sqrt{1-f(r;m)}}{f(r;m)}\,\mathrm{d}r
\fe
removes the coordinate singularity at the outer horizon and gives
\ie
\label{eq:phantom-tunnel-pg-metric}
\mathrm{d}s^{2} =-f(r;m)\mathrm{d}\tau^{2} +2\mathcal{V}(r;m)\mathrm{d}\tau\,\mathrm{d}r +\mathrm{d}r^{2} +\mathcal{R}^{2}(r)\mathrm{d}\Omega^{2},
\fe
with
\ie
\label{eq:phantom-tunnel-shift}
\mathcal{V}(r;m) \equiv\sqrt{1-f(r;m)} =\sqrt{\frac{2m}{r}-\frac{2ma^{2}}{5r^{3}}} =\sqrt{\frac{2m(5r^{2}-a^{2})}{5r^{3}}}.
\fe
The symbol $\mathcal{V}$ denotes the radial shift and must not be confused with the Hamiltonian introduced below.

For an outgoing radial null ray, Eq.~\eqref{eq:phantom-tunnel-pg-metric} leads to
\ie
\label{eq:phantom-tunnel-null-velocity}
\begin{split}
\dot r \equiv\frac{\mathrm{d}r}{\mathrm{d}\tau} &=1-\mathcal{V}(r;m)\\ &=1-\sqrt{\frac{2m}{r}-\frac{2ma^{2}}{5r^{3}}}\\ & \approx \, \, 1-\sqrt{\frac{2m}{r}} +\frac{a^{2}}{10r^{2}}\sqrt{\frac{2m}{r}} +\mathcal{O}\!\left( \frac{a^{4}}{r^{4}}\sqrt{\frac{m}{r}} \right).
\end{split}
\fe
The last term can equivalently be written as $a^{2}\sqrt{m/r^{5}}/(5\sqrt{2})$. The replacement $m=M-\omega'$ must be made before the action is integrated; retaining $M$ in Eq.~\eqref{eq:phantom-tunnel-null-velocity} would remove the backreaction that the tunneling construction is intended to describe.

The instantaneous outer horizon is the perturbative root of $f(r;m)=0$,
\ie
\label{eq:phantom-tunnel-moving-horizon}
r_{\mathrm h}(m) =2m-\frac{a^{2}}{10m} +\mathcal{O}\!\left(\frac{a^{4}}{m^{3}}\right).
\fe
Accordingly, the forbidden trajectory connects
\ie
\label{eq:phantom-tunnel-endpoints}
r_i=r_{\mathrm h}(M), \qquad r_f=r_{\mathrm h}(M-\omega), \qquad r_f<r_i.
\fe
Near the moving horizon, the outgoing velocity has a simple zero:
\ie
\label{eq:phantom-tunnel-near-horizon-velocity}
\dot r =\kappa(m)\bigl[r-r_{\mathrm h}(m)\bigr] +\mathcal{O}\!\left(\bigl[r-r_{\mathrm h}(m)\bigr]^{2}\right),
\fe
where
\ie
\label{eq:phantom-tunnel-running-kappa}
\begin{split}
\kappa(m) &=\frac{1}{2} \left.\frac{\partial f(r;m)}{\partial r}\right|_{r=r_{\mathrm h}(m)} =\frac{1}{4m} \left(1-\frac{a^{2}}{20m^{2}}\right) +\mathcal{O}\!\left(\frac{a^{4}}{m^{5}}\right),\\ \frac{1}{\kappa(m)} &=4m+\frac{a^{2}}{5m} +\mathcal{O}\!\left(\frac{a^{4}}{m^{3}}\right).
\end{split}
\fe

Because the Painlev\'e--Gullstrand time is regular at the horizon, the imaginary contribution in the null geodesic formulation is carried by the radial pole. The relevant part of the classical action is
\ie
\label{eq:phantom-tunnel-action-radial}
\operatorname{Im}\mathcal{S} =\operatorname{Im}\int_{r_i}^{r_f}p_r\,\mathrm{d}r =\operatorname{Im}\int_{r_i}^{r_f} \int_{0}^{p_r}\mathrm{d}p'_r\,\mathrm{d}r.
\fe
Hamilton's equation, $\dot r=(\partial\mathcal{H}/\partial p_r)_r$, gives $\mathrm{d}\mathcal{H}=\dot r\,\mathrm{d}p_r$. Taking $\mathcal{H}=m=M-\omega'$ and $\mathrm{d}\mathcal{H}=-\mathrm{d}\omega'$ then yields
\ie
\label{eq:phantom-tunnel-action-double-integral}
\begin{split}
\operatorname{Im}\mathcal{S} &=\operatorname{Im}\int_{r_i}^{r_f}\mathrm{d}r \int_{M}^{M-\omega} \frac{\mathrm{d}\mathcal{H}}{\dot r} = \operatorname{Im}\int_{0}^{\omega}\mathrm{d}\omega' \int_{r_f}^{r_i} \frac{\mathrm{d}r}{ \dot r\bigl(r;M-\omega'\bigr)}.
\end{split}
\fe
For each intermediate value of $\omega'$, the pole at $r=r_{\mathrm h}(M-\omega')$ lies between the two endpoints. The positive energy Feynman prescription gives
\ie
\label{eq:phantom-tunnel-pole-residue}
\operatorname{Im}\int_{r_f}^{r_i} \frac{\mathrm{d}r}{ \dot r\bigl(r;M-\omega'\bigr)} =\frac{\pi}{\kappa(M-\omega')}.
\fe
Inserting Eq.~\eqref{eq:phantom-tunnel-running-kappa} and retaining terms through $a^{2}$ produces
\ie
\label{eq:phantom-tunnel-imaginary-action}
\begin{aligned}
\operatorname{Im}\mathcal{S}(M,a;\omega) &=\pi\int_{0}^{\omega} \left[ 4(M-\omega')+\frac{a^{2}}{5(M-\omega')} \right]\mathrm{d}\omega'\\ &=4\pi\omega\left(M-\frac{\omega}{2}\right) +\frac{\pi a^{2}}{5} \ln\!\left(\frac{M}{M-\omega}\right).
\end{aligned}
\fe

The semiclassical tunneling probability is therefore
\ie
\label{eq:phantom-tunnel-rate}
\begin{aligned}
\Gamma(\omega;M,a) &\sim\exp\!\left[-2\operatorname{Im}\mathcal{S}(M,a;\omega)\right]\\ &=\exp\!\left[ -8\pi\omega\left(M-\frac{\omega}{2}\right) -\frac{2\pi a^{2}}{5} \ln\!\left(\frac{M}{M-\omega}\right) \right]\\ &=\exp\!\left[-8\pi\omega \left(M-\frac{\omega}{2}\right)\right] \left(\frac{M-\omega}{M}\right)^{2\pi a^{2}/5}.
\end{aligned}
\fe
This expression is exponentially suppressed for $0<\omega<M$. In the limit $a\to0$, it reduces to the standard energy conserving Schwarzschild result
\ie
\label{eq:phantom-tunnel-schwarzschild-limit}
\Gamma_{\mathrm{Schw}}(\omega)
\sim\exp\!\left[-8\pi\omega
\left(M-\frac{\omega}{2}\right)\right].
\fe
The term proportional to $\omega^{2}$ already makes the Schwarzschild tunneling spectrum nonthermal. The phantom DBI geometry adds the logarithmic correction in Eq.~\eqref{eq:phantom-tunnel-rate}.

Identifying $\Gamma$ with the emission to absorption ratio and imposing Bose statistics gives the effective occupation factor
\ie
\label{eq:phantom-tunnel-occupation}
n(\omega;M,a) \equiv\frac{1}{ \exp\!\left[ 8\pi\omega\left(M-\frac{\omega}{2}\right) +\frac{2\pi a^{2}}{5} \ln\!\left(\frac{M}{M-\omega}\right) \right]-1}.
\fe
The corresponding spectral element, before greybody transmission is included, is
\ie
\label{eq:phantom-tunnel-spectral-element}
\mathrm{d}\mathcal{N}^{(\mathrm{tun})}_{\omega} =\frac{\mathrm{d}\omega}{2\pi}\, n(\omega;M,a).
\fe
Because the exponent is nonlinear in $\omega$, this expression is not an exact Planck distribution. It is an effective single emission spectrum; successive quanta are correlated through the changing mass and cannot be represented by independent thermal trials.

The low energy expansion makes contact with the fixed background result:
\ie
\label{eq:phantom-tunnel-low-energy-expansion}
\begin{split}
2\operatorname{Im}\mathcal{S} =&\left(8\pi M+\frac{2\pi a^{2}}{5M}\right)\omega +\left(-4\pi+\frac{\pi a^{2}}{5M^{2}}\right)\omega^{2} +\mathcal{O}\!\left(\frac{a^{2}\omega^{3}}{M^{3}}, \frac{a^{4}}{M^{2}}\right).
\end{split}
\fe
The coefficient linear in $\omega$ determines
\ie
\label{eq:phantom-tunnel-effective-temperature}
T_{\mathrm H} =\left(8\pi M+\frac{2\pi a^{2}}{5M}\right)^{-1} =\frac{1}{8\pi M} \left(1-\frac{a^{2}}{20M^{2}}\right) +\mathcal{O}\!\left(\frac{a^{4}}{M^{5}}\right),
\fe
in agreement with Eq.~\eqref{eq:phantom-temperature-explicit}. The terms of order $\omega^{2}$ and higher retain the recoil information and cannot be absorbed into a constant temperature.

\begin{figure}[!b]
    \centering
    \includegraphics[width=0.6\linewidth]{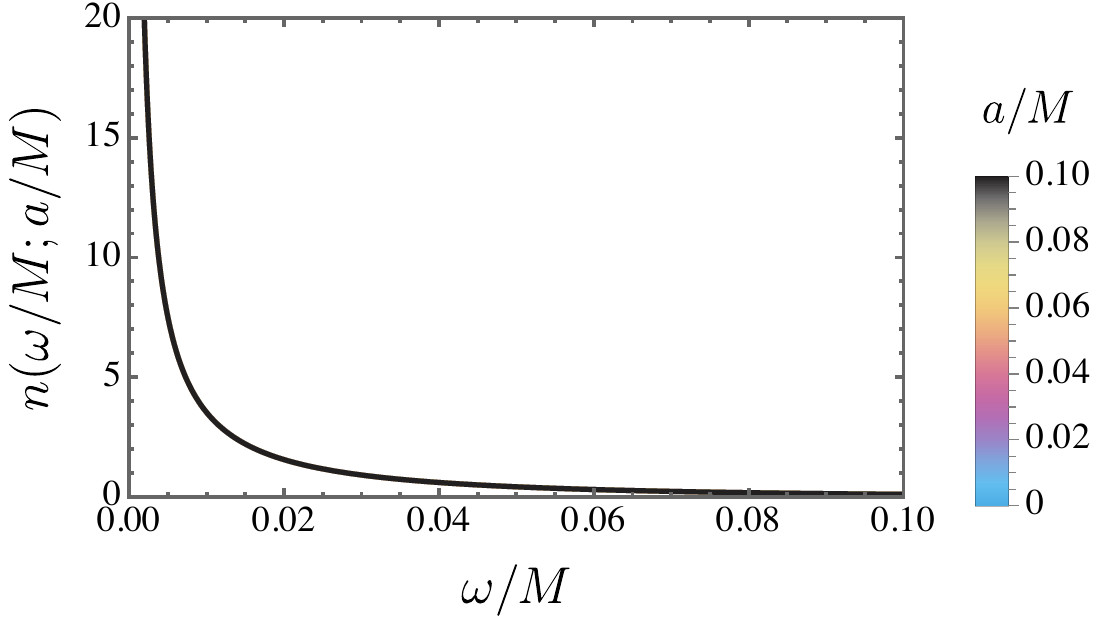}
    \caption{Effective bosonic occupation number $n_{\mathrm{B}}^{\mathrm{tun}}$ obtained from the energy-- conserving tunneling rate as a function of $\omega/M$. The black hole mass was set to $M=1$, while the color scale denotes the ratio $a/M$. }
    \label{fig:phantom-tunnel-bosons}
\end{figure}

For fixed $M$ and $0<\omega<M$, the logarithm in Eq.~\eqref{eq:phantom-tunnel-occupation} is positive. Increasing $a$ therefore raises the tunneling exponent and reduces the bosonic occupation number. Figure~\ref{fig:phantom-tunnel-bosons} displays this suppression for $M=0.2$. The inset isolates the relative change with respect to the Schwarzschild curve, since the perturbative values of $a$ produce closely spaced spectra.

\subsection{Fermionic modes}

The scalar calculation has separated two aspects of the Hawking process. On a fixed background, the analytic continuation of the outgoing modes determines the Bogoliubov mixing and the horizon temperature. Once energy conservation is enforced, the mass varies during the emission and the spectrum acquires nonthermal corrections. The same distinction is required for a Dirac field. The geometric exponential is independent of the spin of the emitted particle, whereas the normalization of the quantum state follows the fermionic anticommutation relations~\cite{o69,o75,o71,o70,o73,o74,o72, vanzo2011tunnelling}.

We consider a neutral Dirac field of mass $m_{\psi}$. Its equation on the phantom Dirac--Born--Infeld background is
\ie
\label{eq:phantom-dirac-equation}
\left(\gamma^{\mu}\nabla_{\mu} +\frac{m_{\psi}}{\hbar}\right)\Psi=0, \qquad \nabla_{\mu} =\partial_{\mu} +\frac{1}{4}\omega_{\mu\hat a\hat b} \gamma^{\hat a}\gamma^{\hat b},
\fe
where $\omega_{\mu\hat a\hat b}$ denotes the spin connection and hatted indices refer to an orthonormal frame. The curved matrices satisfy
\ie
\label{eq:phantom-dirac-clifford}
\left\{\gamma^{\mu},\gamma^{\nu}\right\} =2g^{\mu\nu}\mathbb{1}_{4}.
\fe
For the static metric in Eq.~\eqref{eq:phantom-metric-particle-creation}, a convenient exterior coframe is
\ie
\label{eq:phantom-dirac-static-coframe}
e^{\hat 0}=\sqrt{f(r)}\,\mathrm{d}t, \qquad e^{\hat 1}=\frac{\mathrm{d}r}{\sqrt{f(r)}}, \qquad e^{\hat 2}=\mathcal{R}(r)\,\mathrm{d}\theta, \qquad e^{\hat 3}=\mathcal{R}(r)\sin\theta\,\mathrm{d}\phi.
\fe
This distinction is required because the radial coordinate of the phantom geometry is not the areal radius.

Writing the spinor in the eikonal form
\ie
\label{eq:phantom-dirac-wkb-ansatz}
\Psi =\exp\!\left(\frac{i}{\hbar}\mathcal{I}\right) \left(\mathsf{u}_{0}+\hbar\mathsf{u}_{1}+\cdots\right),
\fe
the leading contribution to Eq.~\eqref{eq:phantom-dirac-equation} gives
\ie
\label{eq:phantom-dirac-leading-equation}
\left(i\gamma^{\mu}\partial_{\mu}\mathcal{I}
+m_{\psi}\right)\mathsf{u}_{0}=0.
\fe
A nontrivial spinor amplitude requires
\ie
\label{eq:phantom-dirac-hamilton-jacobi}
g^{\mu\nu} \partial_{\mu}\mathcal{I}\, \partial_{\nu}\mathcal{I} +m_{\psi}^{2}=0.
\fe
The spin connection first enters the transport equation for $\mathsf{u}_{0}$; it does not alter the singular part of the eikonal phase. This observation is sufficient for both calculations below. We restrict the explicit formulas to $m_{\psi}=0$, which is the regime relevant to Hawking emission when the temperature is well above the particle mass.


\subsubsection{Bogoliubov description on the fixed background}

The massless phase may be separated as $\mathcal{I}=-\omega t+W(r)+J(\theta,\phi)$. The radial part of Eq.~\eqref{eq:phantom-dirac-hamilton-jacobi} satisfies
\ie
\label{eq:phantom-dirac-static-radial-action}
W'_{\pm}(r)=\pm\frac{\omega}{f(r)}.
\fe
The two signs reproduce the outgoing and ingoing phases $e^{-i\omega u}$ and $e^{-i\omega v}$, respectively. The angular function and the spinor transported along the ray remain regular at the outer horizon and do not modify the logarithmic relation between $u$ and $v$.

Let $F_{\omega\lambda}$ and $F^{c}_{\omega\lambda}$ denote positive frequency particle and antiparticle modes on past null infinity. The label $\lambda$ collects the angular and spin quantum numbers, while the superscript $c$ denotes charge conjugation. The same field may be expanded in an exterior basis $P_{\omega\lambda}$ and in a complementary basis $Q_{\omega\lambda}$ with no outgoing support on future null infinity:
\ie
\label{eq:phantom-dirac-field-bases}
\begin{split}
\Psi =&\sum_{\lambda}\int_{0}^{\infty}\mathrm{d}\omega \left( F_{\omega\lambda}a_{\omega\lambda} +F^{c}_{\omega\lambda}b^{\dagger}_{\omega\lambda} \right)\\ =&\sum_{\lambda}\int_{0}^{\infty}\mathrm{d}\omega \left( P_{\omega\lambda}c_{\omega\lambda} +P^{c}_{\omega\lambda}d^{\dagger}_{\omega\lambda} +Q_{\omega\lambda}\widetilde c_{\omega\lambda} +Q^{c}_{\omega\lambda}\widetilde d^{\dagger}_{\omega\lambda} \right).
\end{split}
\fe
All creation and annihilation operators obey canonical anticommutation relations. The two spin polarizations are degenerate because the background is static and spherically symmetric.

The ray tracing relation derived in Eq.~\eqref{eq:phantom-ray-tracing-explicit} applies without change to the spinorial phase. With
\ie
\label{eq:phantom-dirac-xi}
\Xi(M,a) =4M+\frac{a^{2}}{5M} =\frac{1}{\kappa_{\mathrm h}} +\mathcal{O}\!\left(\frac{a^{4}}{M^{3}}\right),
\fe
an exterior mode traced back to past null infinity behaves as
\ie
\label{eq:phantom-dirac-traced-mode}
P_{\omega\lambda}(v) =K_{\omega\lambda}\, \Theta(v_{0}-v) \left(\frac{v_{0}-v}{C_{v}}\right)^{i\omega\Xi(M,a)} \mathsf{u}_{\lambda}(v).
\fe
Here $\mathsf{u}_{\lambda}(v)$ is a spinor that is regular at $v=v_{0}$. Its value at the last escaping ray, together with the chosen Dirac normalization, is absorbed into $K_{\omega\lambda}$.

The mode decomposition
\ie
\label{eq:phantom-dirac-bogoliubov-expansion}
P_{\omega\lambda} =\int_{0}^{\infty}\mathrm{d}\omega' \left[ \alpha^{(\mathrm F,\lambda)}_{\omega\omega'} F_{\omega'\lambda} +\beta^{(\mathrm F,\lambda)}_{\omega\omega'} F^{c}_{\omega'\bar\lambda} \right]
\fe
mixes the particle mode with the charge conjugate mode carrying the paired quantum numbers $\bar\lambda$. In general lines, the singular power in Eq.~\eqref{eq:phantom-dirac-traced-mode} produces the same features encountered in the scalar calculation. Analyticity, in the lower half of the complex, $v$ plane fixes the relative amplitude of its positive and negative frequency parts. The regular spinor transport and the normalization of the angular harmonics multiply both parts by the same modulus and do not enter the quotient, as we should expect. For late time wave packets, we find
\ie
\label{eq:phantom-dirac-bogoliubov-ratio}
\frac{ \left|\alpha^{(\mathrm F,\lambda)}_{\omega}\right|^{2} }{ \left|\beta^{(\mathrm F,\lambda)}_{\omega}\right|^{2} } =\exp\!\left[ 2\pi\omega\Xi(M,a) \right] =\exp\!\left[ \left(8\pi M+\frac{2\pi a^{2}}{5M}\right)\omega \right].
\fe
In a similar manner, $|\beta^{(\mathrm F,\lambda)}_{\omega}|/ |\alpha^{(\mathrm F,\lambda)}_{\omega}| =\exp[-\pi\omega\Xi(M,a)]$. This relation is geometric and coincides with the bosonic quotient. The difference appears when the transformation is normalized.

After diagonalizing the wave packet basis, the regular and horizon adapted fermionic operators are related by an $SU(2)$ transformation,
\ie
\label{eq:phantom-dirac-operator-transformation}
\begin{pmatrix}
c^{(\mathrm K)}_{\omega\lambda}\\ d^{(\mathrm K)\dagger}_{\omega\bar\lambda} \end{pmatrix} = \begin{pmatrix} \widehat{\alpha}^{(\mathrm F)}_{\omega} & -\widehat{\beta}^{(\mathrm F)}_{\omega}\\ \widehat{\beta}^{(\mathrm F)*}_{\omega} & \widehat{\alpha}^{(\mathrm F)*}_{\omega} \end{pmatrix} \begin{pmatrix} c^{(\mathrm I)}_{\omega\lambda}\\ d^{(\mathrm{II})\dagger}_{\omega\bar\lambda}
\end{pmatrix},
\fe
where regions $\mathrm I$ and $\mathrm{II}$ denote the exterior and interior sectors, respectively, and the superscript $\mathrm K$ identifies modes that are analytic across the future horizon. Up to convention dependent phases, the normalized Bogoliubov coefficients are explicitly
\ie
\label{eq:phantom-dirac-normalized-coefficients}
\begin{aligned}
\widehat{\alpha}^{(\mathrm F)}_{\omega}(M,a) &=\frac{e^{i\delta_{\alpha}}}{ \sqrt{1+\exp\!\left[- \left(8\pi M+\frac{2\pi a^{2}}{5M}\right)\omega \right]}},\\ \widehat{\beta}^{(\mathrm F)}_{\omega}(M,a) &=\frac{e^{i\delta_{\beta}}}{ \sqrt{1+\exp\!\left[ \left(8\pi M+\frac{2\pi a^{2}}{5M}\right)\omega \right]}}.
\end{aligned}
\fe
They satisfy
\ie
\label{eq:phantom-dirac-fermionic-normalization}
\left|\widehat{\alpha}^{(\mathrm F)}_{\omega}\right|^{2} +\left|\widehat{\beta}^{(\mathrm F)}_{\omega}\right|^{2}=1, \qquad \frac{\left|\widehat{\alpha}^{(\mathrm F)}_{\omega}\right|^{2}} {\left|\widehat{\beta}^{(\mathrm F)}_{\omega}\right|^{2}} =\exp\!\left[ \left(8\pi M+\frac{2\pi a^{2}}{5M}\right)\omega \right].
\fe
The plus sign in the first relation is fixed by the fermionic anticommutation algebra. It replaces the difference of squared coefficients that appears for a bosonic field.

The occupation number of exterior particles, per spin and angular mode, is
\ie
\label{eq:phantom-dirac-fixed-occupation}
n^{(\mathrm H)}_{\mathrm F}(\omega;M,a) =\left|\widehat{\beta}^{(\mathrm F)}_{\omega}\right|^{2} =\frac{1}{ \exp\!\left[ \left(8\pi M+\frac{2\pi a^{2}}{5M}\right)\omega \right]+1}.
\fe
It has the Fermi--Dirac form at the temperature
\ie
\label{eq:phantom-dirac-temperature}
T_{\mathrm H} =\frac{1}{8\pi M} \left(1-\frac{a^{2}}{20M^{2}}\right) +\mathcal{O}\!\left(\frac{a^{4}}{M^{5}}\right),
\fe
which is the same temperature found from the scalar field. At fixed $M$, the parameter $a$ lowers the temperature and suppresses every mode with $\omega>0$. The two spin polarizations give the same contribution. The flux measured at infinity also contains the fermionic transmission probabilities:
\ie
\label{eq:phantom-dirac-fixed-flux}
\frac{\mathrm{d}^{2}\mathcal{N}^{(\mathrm F)}_{\infty}} {\mathrm{d}t\,\mathrm{d}\omega} =\frac{1}{2\pi} \sum_{\lambda} \frac{\mathscr{T}^{(\mathrm F)}_{\omega\lambda}} {\exp(\omega/T_{\mathrm H})+1}.
\fe
The coefficients $\mathscr{T}^{(\mathrm F)}_{\omega\lambda}$ describe scattering outside the horizon and do not alter the Bogoliubov quotient.


\subsubsection{Energy-conserving fermionic tunneling}

The fixed background spectrum does not include the decrease of the black hole mass. To incorporate this effect, we employ the same instantaneous mass $m=M-\omega'$ introduced in Eq.~\eqref{eq:phantom-tunnel-running-mass}. One additional point is important to mention: the symbol $m$ refers here to the black hole mass along the trajectory and must not be confused with the field mass $m_{\psi}$, which has been set to zero.

In Painlev\'e--Gullstrand coordinates, the metric may be written as
\ie
\label{eq:phantom-dirac-pg-metric}
\begin{split}
\mathrm{d}s^{2} & = -f(r;m)\mathrm{d}\tau^{2} +2\mathcal{V}(r;m)\mathrm{d}\tau\,\mathrm{d}r +\mathrm{d}r^{2} +\mathcal{R}^{2}(r)\mathrm{d}\Omega^{2}\\ &=-\mathrm{d}\tau^{2} +\left[\mathrm{d}r +\mathcal{V}(r;m)\mathrm{d}\tau\right]^{2} +\mathcal{R}^{2}(r)\mathrm{d}\Omega^{2},
\end{split}
\fe
where
\ie
\nonumber
\label{eq:phantom-dirac-pg-functions}
f(r;m) =1-\frac{2m}{r}+\frac{2ma^{2}}{5r^{3}}, \qquad \mathcal{V}(r;m) =\sqrt{1-f(r;m)}.
\fe
The second line of Eq.~\eqref{eq:phantom-dirac-pg-metric} provides the regular
coframe
\ie
\label{eq:phantom-dirac-pg-coframe}
e^{\hat 0}=\mathrm{d}\tau, \qquad e^{\hat 1}=\mathrm{d}r +\mathcal{V}(r;m)\mathrm{d}\tau, \qquad e^{\hat 2}=\mathcal{R}(r)\mathrm{d}\theta, \qquad e^{\hat 3}=\mathcal{R}(r)\sin\theta\mathrm{d}\phi.
\fe
No tetrad component diverges at the moving horizon.

The inverse radial metric has $g^{\tau\tau}=-1$, $g^{\tau r}=\mathcal{V}$, and $g^{rr}=f$. For a radial phase $\mathcal{I}=-\omega\tau+W(r)$, Eq.~\eqref{eq:phantom-dirac-hamilton-jacobi} becomes
\ie
\label{eq:phantom-dirac-pg-hj}
-\omega^{2} -2\omega\mathcal{V}(r;m)W'(r) +f(r;m)W'(r)^{2}=0. \fe Its two branches are \ie \label{eq:phantom-dirac-pg-momenta} \begin{split} W'_{\mathrm{out}}(r) &=\frac{\omega[1+\mathcal{V}(r;m)]}{f(r;m)} =\frac{\omega}{1-\mathcal{V}(r;m)},\\ W'_{\mathrm{in}}(r) &=-\frac{\omega[1-\mathcal{V}(r;m)]}{f(r;m)} =-\frac{\omega}{1+\mathcal{V}(r;m)}.
\end{split}
\fe
The ingoing momentum is regular at $r=r_{\mathrm h}(m)$, where $\mathcal{V}=1$. The outgoing momentum contains the simple pole
\ie
\label{eq:phantom-dirac-pg-pole}
W'_{\mathrm{out}}(r) =\frac{\omega}{ \kappa(m)[r-r_{\mathrm h}(m)]} +\text{terms regular at }r=r_{\mathrm h}(m).
\fe
The spin connection affects the transport of the polarization through the regular terms, but it does not change this residue. Both spin states therefore have the same leading tunneling factor.

Energy conservation requires the pole to move while the quantum crosses the horizon. Combining Eq.~\eqref{eq:phantom-dirac-pg-pole} with Hamilton's equation gives
\ie
\label{eq:phantom-dirac-backreacted-action}
\begin{split}
\operatorname{Im}\mathcal{S}_{\mathrm F}(M,a;\omega) &=\pi\int_{0}^{\omega} \frac{\mathrm{d}\omega'}{\kappa(M-\omega')} = 4\pi\omega\left(M-\frac{\omega}{2}\right) +\frac{\pi a^{2}}{5} \ln\!\left(\frac{M}{M-\omega}\right),
\end{split}
\fe
where terms beyond $a^{2}$ have been discarded. The same imaginary action was obtained for the scalar shell because the leading eikonal equation is spin independent. The fermionic tunneling probability is
\ie
\label{eq:phantom-dirac-backreacted-rate}
\begin{aligned}
\Gamma_{\mathrm F}(\omega;M,a) &\sim\exp\!\left[-2 \operatorname{Im}\mathcal{S}_{\mathrm F}(M,a;\omega)\right]\\ &=\exp\!\left[ -8\pi\omega\left(M-\frac{\omega}{2}\right) -\frac{2\pi a^{2}}{5} \ln\!\left(\frac{M}{M-\omega}\right) \right]\\ &=\exp\!\left[-8\pi\omega \left(M-\frac{\omega}{2}\right)\right] \left(\frac{M-\omega}{M}\right)^{2\pi a^{2}/5}.
\end{aligned}
\fe
The result is valid for $0<\omega<M$ together with $a/(M-\omega)\ll1$. In the limit $a\to0$, it becomes the energy conserving Schwarzschild factor $\exp[-8\pi\omega(M-\omega/2)]$.

Using $\Gamma_{\mathrm F}$ as the emission to absorption ratio, fermionic normalization gives the effective occupation number
\ie
\label{eq:phantom-dirac-tunneling-occupation}
n^{(\mathrm{tun})}_{\mathrm F}(\omega;M,a) =\frac{\Gamma_{\mathrm F}}{1+\Gamma_{\mathrm F}} =\frac{1}{ \exp\!\left[ 8\pi\omega\left(M-\frac{\omega}{2}\right) +\frac{2\pi a^{2}}{5} \ln\!\left(\frac{M}{M-\omega}\right) \right]+1}.
\fe
This expression differs from the bosonic result only through the sign in the statistical denominator. Notice that it is not exactly thermal because the exponent contains the recoil term and the logarithmic DBI correction. In the low--energy regime,
\ie
\label{eq:phantom-dirac-tunneling-low-energy}
n^{(\mathrm{tun})}_{\mathrm F} =\frac{1}{\exp(\omega/T_{\mathrm H})+1} +\mathcal{O}(\omega^{2}),
\fe
with the temperature given by Eq.~\eqref{eq:phantom-dirac-temperature}. At $\omega\to0$, the occupation tends to $1/2$ per spin mode, rather than diverging as in the bosonic sector. This behavior is the direct consequence of Pauli exclusion.

\begin{figure}[!b]
    \centering
    \includegraphics[width=0.6\linewidth]{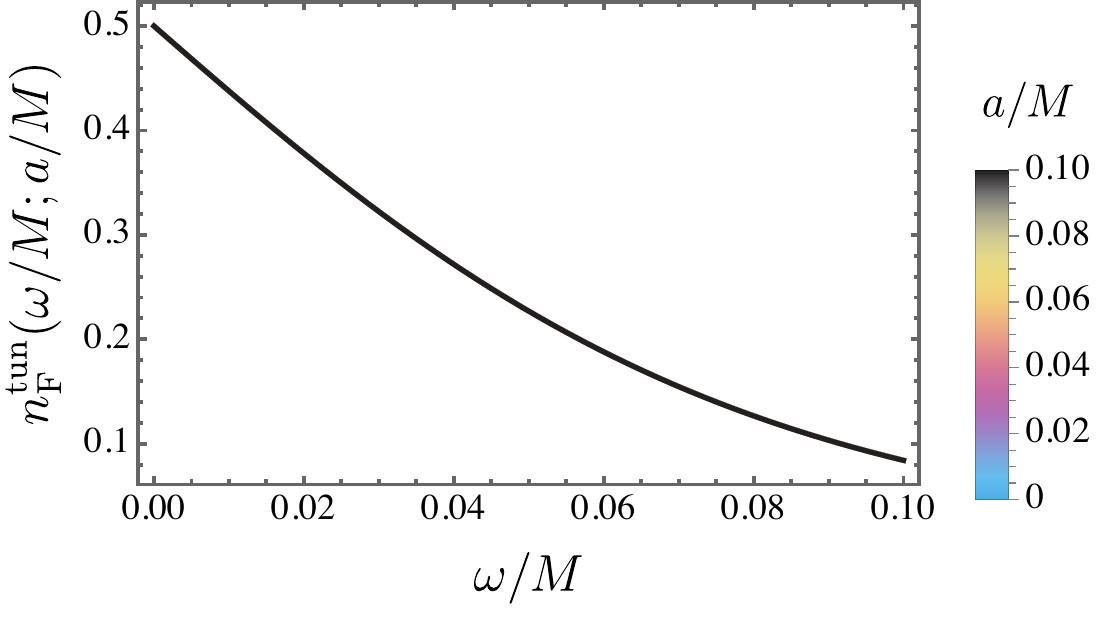}
    \caption{Effective fermionic occupation number $n_{\mathrm{F}}^{\mathrm{tun}}$ per spin mode obtained from the energy--conserving tunneling rate as a function of $\omega/M$. The black hole mass was set to $M=1$, while the color scale denotes the ratio $a/M$.}
    \label{fig:phantom-dirac-tunnel}
\end{figure}

For fixed $M$ and $0<\omega<M$, the DBI contribution to the exponent in Eq.~\eqref{eq:phantom-dirac-tunneling-occupation} is positive. Increasing $a$ therefore suppresses fermionic emission. Figure~\ref{fig:phantom-dirac-tunnel} shows this effect for $M=0.2$ and the same parameter choices employed in the bosonic plot. The inset displays the relative suppression with respect to $a=0$.

\section{Evaporation rate}
\label{sec:evaporation}

The occupation numbers derived in the preceding section describe particle creation mode by mode.  The evolution of the ADM mass is governed instead by the total energy flux that reaches future null infinity.  For independent horizon modes, the statistical partition function may be written formally as
\ie
\mathcal{Z}_{\rm H} =\prod_{i,\Lambda,\omega} \left(1-\eta_i e^{-\omega/T_{\rm H}}\right)^{-\eta_i}, \qquad \eta_i=(-1)^{2s_i}.
\label{partitionevap}
\fe
The product in Eq.~\eqref{partitionevap} belongs to the state counting problem.  Since observable fluxes follow from derivatives of $\ln\mathcal{Z}_{\rm H}$, the contributions of different species and angular channels add; they must not be multiplied.  For the neutral and nonrotating geometry considered here, the corresponding luminosity is
\ie
P_{\infty}(M,a)\equiv-\frac{\mathrm{d}M}{\mathrm{d}t} =\sum_i\frac{g_i}{2\pi} \sum_{\Lambda}d_{\Lambda}^{(i)} \int_{m_i}^{\infty} \frac{\omega\,\mathcal{T}_{\Lambda}^{(i)}(\omega;M,a)} {\exp(\omega/T_{\rm H})-\eta_i}\,\mathrm{d}\omega .
\label{fullpower}
\fe
Here, $i$ labels the emitted species, $g_i$ counts its nonangular internal states, $\Lambda$ denotes a complete angular channel, and $d_{\Lambda}^{(i)}$ is its multiplicity.  The transmission probability $\mathcal{T}_{\Lambda}^{(i)}$ contains the greybody filtering between the horizon and infinity.  Eq.~\eqref{fullpower}, rather than an uncorrected Stefan--Boltzmann law, is the appropriate starting point whenever the complete evaporation history is required \cite{page1976particle,hiscock1990evolution}.

It is useful to express the same result through the absorption cross section. Writing $k_i(\omega)=\sqrt{\omega^2-m_i^2}$ and defining the cross section per internal state by
\ie
\sigma_i(\omega;M,a) =\frac{\pi}{k_i^2} \sum_{\Lambda}d_{\Lambda}^{(i)} \mathcal{T}_{\Lambda}^{(i)}(\omega;M,a),
\fe
we have
\ie
P_{\infty}(M,a) =\sum_i\frac{g_i}{2\pi^2} \int_{m_i}^{\infty} \frac{\omega k_i^2\sigma_i(\omega;M,a)} {\exp(\omega/T_{\rm H})-\eta_i}\,\mathrm{d}\omega .
\label{crosspower}
\fe
This form keeps the particle content and the propagation through the exterior potential separate. It also shows why replacing all greybody effects by one constant coefficient generally loses information: the transmission depends on the species, the angular channel, the frequency, and the instantaneous values of $M$ and $a$.

In the geometric--optics regime, $\omega b_{\rm c}\gg1$, the absorption cross section approaches the capture value $\sigma_{\rm geo}=\pi b_{\rm c}^{2}$. More precisely, the partial transmissions approach a capture step in angular momentum, while their sum tends to $\sigma_{\rm geo}$; the individual greybody factors do not become identically equal to unity for every channel. If the thermally accessible particles are effectively massless, the frequency integral in Eq.~\eqref{crosspower} gives
\ie
P_{\rm geo}(M,a) =\frac{\pi^2}{30}\,g_{*}(T_{\rm H})\, \sigma_{\rm geo}(M,a)T_{\rm H}^{4}, \qquad g_{*}=\sum_{i\in{\rm bosons}}g_i +\frac{7}{8}\sum_{i\in{\rm fermions}}g_i.
\label{geopower}
\fe
Only the case where satisfying $m_i\ll T_{\rm H}$ should be retained in $g_{*}$. Mass thresholds make $g_{*}$ piecewise dependent on $M$, even before the frequency dependence of the greybody factors is restored.

For the exact phantom Dirac--Born--Infeld geometry, the unstable null orbit is located at $r_{\rm ph}=3M$.  Introducing only in this section the dimensionless ratio $\zeta=a/(3M)$, its critical impact parameter and geometric cross section are
\ie
b_{\rm c} =3M\left(\frac{\zeta^3}{\zeta-\tan^{-1}\zeta}\right)^{1/2}, \qquad \sigma_{\rm geo} =9\pi M^2\frac{\zeta^3}{\zeta-\tan^{-1}\zeta}.
\label{phantomgeocross}
\fe
These relations follow from the unexpanded metric. In the small core domain, they reduce to
\ie
\begin{aligned}
b_{\rm c}&=3\sqrt{3}M \left[1+\frac{a^2}{30M^2} +\mathcal{O}\!\left(\frac{a^4}{M^4}\right)\right],\\ \sigma_{\rm geo}&=27\pi M^2 \left[1+\frac{a^2}{15M^2} +\mathcal{O}\!\left(\frac{a^4}{M^4}\right)\right].
\end{aligned}
\label{phantomgeocrossexp}
\fe
The coordinate position of the photon sphere therefore has no correction at order $a^2$; the increase in the capture cross section originates from the areal radius and from the value of the lapse at the null orbit.

Combining Eq.~\eqref{phantomgeocrossexp} with the Hawking temperature obtained previously gives
\ie
\sigma_{\rm geo}T_{\rm H}^{4} =\frac{27}{4096\pi^3M^2} \left[1-\frac{2a^2}{15M^2} +\mathcal{O}\!\left(\frac{a^4}{M^4}\right)\right].
\fe
In other words, the high--frequency mass evolution for a fixed set of massless degrees of freedom is
\ie
\frac{\mathrm{d}M}{\mathrm{d}t} =-\frac{9g_{*}}{40960\pi M^2} \left[1-\frac{2a^2}{15M^2} +\mathcal{O}\!\left(\frac{a^4}{M^4}\right)\right].
\label{phantommassloss}
\fe
At fixed $M\gg a$, the regular core enlarges the capture cross section but lowers $T_{\rm H}$.  The fourth power of the temperature supplies the dominant correction, so the net luminosity is smaller than its Schwarzschild geometric--optics value.

A particularly transparent illustration is obtained by retaining only a massless spin--$1$ field.  Its two helicities give $g_{\gamma}=2$, and Eq.~\eqref{phantommassloss} becomes
\ie
\left.\frac{\mathrm{d}M}{\mathrm{d}t}\right|_{\gamma} =-\frac{9}{20480\pi M^2} \left[1-\frac{2a^2}{15M^2} +\mathcal{O}\!\left(\frac{a^4}{M^4}\right)\right].
\label{photonmassloss}
\fe
This equation describes the electromagnetic channel, not the full luminosity, as we have argued before. Scalar, fermionic, vector, and gravitational degrees of freedom must be added through Eq.~\eqref{fullpower} when they are kinematically accessible. The photon--only expression is nevertheless useful because the leading eikonal capture cross section is independent of the spin of a test field; spin enters through subleading greybody structure.

If $g_{*}$ remains constant and both limits satisfy $M_i>M_f\gg a$, integration of Eq.~\eqref{phantommassloss} yields
\ie
t_{\rm geo}(M_i\!\to M_f) =\frac{40960\pi}{27g_{*}} \left[ M_i^3-M_f^3+\frac{2a^2}{5}(M_i-M_f) \right] +\mathcal{O}\!\left(\frac{a^4}{M_f}\right).
\label{geolifetime}
\fe
For the spin--$1$ channel alone,
\ie
t_{\gamma}(M_i\!\to M_f) =\frac{20480\pi}{27} \left[ M_i^3-M_f^3+\frac{2a^2}{5}(M_i-M_f) \right] +\mathcal{O}\!\left(\frac{a^4}{M_f}\right).
\label{photonlifetime}
\fe
The positive term proportional to $a^2$ is the lifetime extension predicted by the controlled small core expansion. Again, one point must be emphitized again: the numerical coefficient in Eq.~\eqref{photonlifetime} belongs to the geometric--optics approximation and must not be identified with the Page lifetime obtained from the complete electromagnetic greybody spectrum.

The full geometry permits a broader formulation without expanding in $a/M$. Let $y=r_h/a$.  The exact horizon equation and the standard surface gravity give the parametric relations
\ie
\begin{aligned}
M(y) &=\frac{a}{3\left[(1+y^2)\tan^{-1}(y^{-1})-y\right]},\\ T_{\rm H}(y) &=\frac{1-y\tan^{-1}(y^{-1})} {2\pi a\left[(1+y^2)\tan^{-1}(y^{-1})-y\right]}.
\end{aligned}
\label{exacthorizonparam}
\fe
Together with Eq.~\eqref{phantomgeocross}, these expressions provide an unexpanded geometric--optics model.  For a general particle spectrum, however, the evaporation time is most safely left in the form
\ie
t(M_i\!\to M_f) =\int_{M_f}^{M_i}\frac{\mathrm{d}M}{P_{\infty}(M,a)},
\label{exactlifetime}
\fe
where $P_{\infty}$ is evaluated from Eq.~\eqref{fullpower}. The integral is normally numerical because the greybody factors evolve with the mass and new particle species enter when $T_{\rm H}$ crosses their mass thresholds. In addition, recent development in Hawking radiation has been considered \cite{ahmed2026hawking}.

The endpoint requires particular care.  From Eq.~\eqref{exacthorizonparam},
\ie
\lim_{y\to0^{+}}M(y)=\frac{2a}{3\pi}, \qquad \lim_{y\to0^{+}}T_{\rm H}(y)=\frac{1}{\pi^2a}.
\label{endpointlimits}
\fe
The first limit locates the lower mass boundary of the positive radius black hole. The second shows that this boundary is not selected by $T_{\rm H}=0$ when the temperature is computed from the exact lapse. A stable relic may still follow from the global completion of the geometry or from physics beyond the semiclassical approximation, but it is not established by the Stefan--Boltzmann argument. Likewise, Eqs.~\eqref{geolifetime} and \eqref{photonlifetime} cannot be extrapolated to $M=2a/(3\pi)$ because their derivation assumes $a/M\ll1$.


\section{Partial wave analysis{}}

In this section, we investigate the interaction of a massless scalar wave with the black hole geometry by means of a partial wave analysis. This framework allows us to determine both the absorption and scattering properties of the spacetime. Neglecting the backreaction of the scalar perturbation on the background geometry, the field is treated as a test field satisfying the covariant Klein--Gordon equation
\begin{equation}
\frac{1}{\sqrt{-g}}
\partial_{\mu}\left(
\sqrt{-g}~g^{\mu\nu}\partial_{\nu}\Phi
\right)=0.
\label{eq:KG}
\end{equation}

Owing to the static and spherically symmetric character of the metric described in Eq.~\eqref{eq:DBImetric}--\eqref{eq:lapse}, the temporal, radial, and angular dependence of the field can be separated. We therefore decompose the scalar field as
\begin{equation}
\Phi(t,r,\theta,\phi)
=
\sum_{\ell=0}^{\infty}
\sum_{m=-\ell}^{\ell}
\frac{\psi_{\omega\ell}(r)}{R(r)}
Y_{\ell m}(\theta,\phi),
e^{-i\omega t},
\label{eq:separation}
\end{equation}
where $Y_{\ell m}(\theta,\phi)$ are the standard spherical harmonics, $\omega$ denotes the wave frequency, and $\ell$ and $m$ are the multipole number and the azimuthal number, respectively. Substitution of Eq.~\eqref{eq:separation} into Eq.~\eqref{eq:KG}, followed by the separation of the angular sector, leads to
\begin{equation}
\frac{\mathrm{d}}{\mathrm{d}r}
\left(
f(r)\frac{\mathrm{d}\psi_{\omega\ell}}{\mathrm{d}r}
\right)
+
\left(
\frac{\omega^{2}}{f(r)}
-\frac{\ell(\ell+1)}{R(r)^{2}}
-\frac{f'(r)R'(r)+f(r)R''(r)}{R(r)}
\right)\psi_{\omega\ell}
=0,
\label{eq:radial}
\end{equation}
where a prime denotes differentiation with respect to $r$.
For the analysis of wave propagation, it is convenient to introduce the tortoise coordinate $r^{*}$, defined through
$\frac{\mathrm{d}r^{*}}{\mathrm{d}r}
=
{1}/{f(r)}$. In this coordinate, the event horizon is mapped to $r^{*}\rightarrow-\infty$, while the asymptotically flat region corresponds to $r^{*}\rightarrow+\infty$. The radial equation then assumes the one dimensional Schr\"{o}dinger--like form
\begin{equation}
\frac{\mathrm{d}^{2}\psi_{\omega\ell}}
{\mathrm{d}{r^{*}}^{2}}
+
\left(
\omega^{2}-V_{\text{eff}}(r)
\right)\psi_{\omega\ell}
=0,
\label{eq:radial2}
\end{equation}
with the effective potential
\begin{align}
V_{{\text{eff}}}(r)
&=f(r)\left(\frac{\ell(\ell+1)}{R(r)^{2}}
+\frac{f'(r)R'(r)+f(r)R''(r)}{R(r)}\right)\\
&=f(r) \frac{a^2 f(r)+\left(a^2+r^2\right) \left(r f'(r)+\ell(\ell+1) \right)}{\left(a^2+r^2\right)^2}
\label{eq:potential}.
\end{align}

For an asymptotically flat black hole, the effective potential vanishes both at the event horizon and at spatial infinity, while it generally develops a finite barrier in the intermediate region. Consequently, the radial problem can be interpreted as a standard scattering process in which an incident wave is partially transmitted through the potential barrier and absorbed by the black hole, while the remaining part is reflected toward infinity. This formulation provides the basis for calculating the reflection and transmission coefficients, the partial absorption probabilities, and the corresponding absorption and scattering cross sections.

Since the effective potential in Eq.~\eqref{eq:potential} forms a localized barrier and vanishes both near the event horizon and at spatial infinity, the radial solution reduces to free-wave behavior in these asymptotic regions. For a wave incident from infinity, the physically relevant boundary conditions require a purely ingoing wave at the horizon and a superposition of incident and reflected waves at infinity. Accordingly, we have the following condition~\cite{macedo2015scattering,macedo2016absorption,anacleto2023absorption}
\begin{equation}\label{bound}
	\psi(r^*) \approx 
	\begin{cases}
	{\mathcal{R}_{\omega \ell}} \psi_{{I}}  &\text{for} \ {r^*} \to r_{\rm h} ~ (r \to -\infty )\\
	\psi_{{II}}+\mathcal{T}_{\omega \ell} \psi^*_{{II}} &\text{for} \ {r^*} \to \infty ~ (r \to \infty ).\\
	\end{cases}
\end{equation}
Here, $\mathcal{R}_{\omega\ell}$ and $\mathcal{T}_{\omega\ell}$ denote the reflection and transmission amplitudes, respectively. The asymptotic solutions of Eq.~\eqref{eq:radial2} may be expressed as~\cite{macedo2013absorption}
\begin{align}
\psi_{I}
&=
e^{-i\omega r^*}
\sum_{j=0}^{N}
b_{r_{\rm h}}^{j}(r-r_{\rm h})^{j},
\label{Rroman1}\\
\psi_{II}
&=
e^{-i\omega r^*}
\sum_{j=0}^{N}
\frac{b_{\infty}^{j}}{r^{j}}.
\label{Rroman2}
\end{align}
The coefficients $b_{r_{\rm h}}^{j}$ and $b_{\infty}^{j}$ are determined by substituting these expansions into the radial equation and solving the resulting relations. The numerical integration is then performed subject to the boundary conditions in Eq.~\eqref{bound}, allowing the reflection and transmission amplitudes to be extracted for each frequency and multipole number.

The corresponding reflection and transmission probabilities are given by
$\left|\mathcal{R}_{\omega\ell}\right|^{2}$ and
$\left|\mathcal{T}_{\omega\ell}\right|^{2}$, respectively. Conservation of the radial flux implies
\begin{equation}
\left|\mathcal{R}_{\omega\ell}\right|^{2}
+
\left|\mathcal{T}_{\omega\ell}\right|^{2}
=1.
\end{equation}

The phase shift associated with each partial wave is defined through~\cite{dolan2009scattering}
\begin{equation}
e^{2i\delta_{\omega\ell}}
=
(-1)^{\ell+1}\mathcal{R}_{\omega\ell}.
\label{phase}
\end{equation}
Once the phase shifts and transmission amplitudes are known, the absorption and scattering cross sections can be evaluated.
  
\section{Absorption cross section{}}

\begin{figure}[ht!]
\centering
\includegraphics[width=90mm]{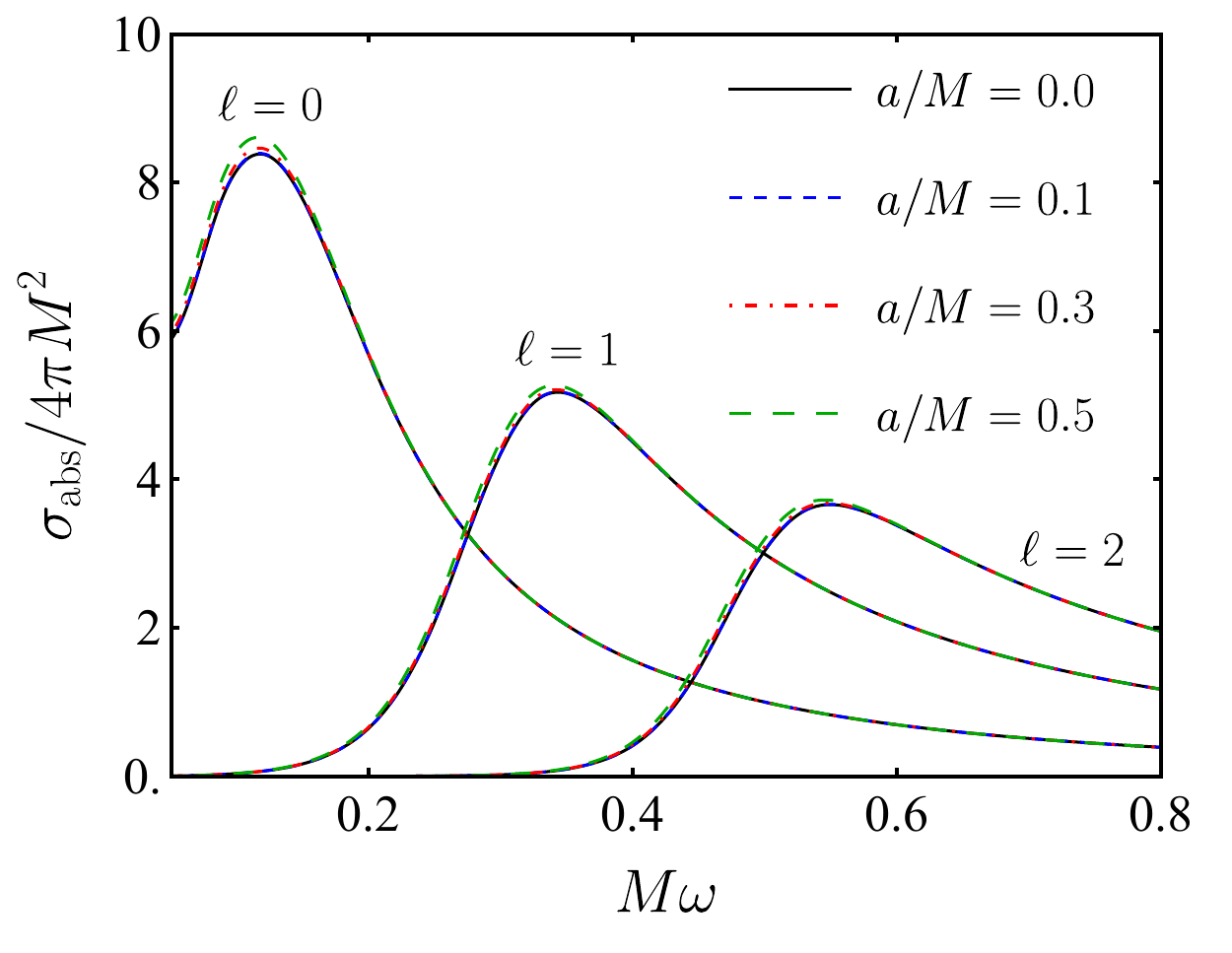}
\caption{Partial absorption cross sections as functions of the dimensionless frequency $M\omega$ for $\ell=0,1,2$ and selected values of the normalized regular core scale $a/M$.}
\label{fig:Psigma}
\end{figure}

The total absorption cross section is obtained by summing the partial contributions using the partial wave technique \cite{futterman1986scattering,Marco1,Marco2,Marco3}, \begin{equation} \label{abs_crosssection} \sigma_{\rm abs} = \sum_{\ell = 0}^{\infty} \sigma_{\rm abs}^{\ell}, 
\end{equation} 
where the partial cross section depends on the phase shift parameter defined in Eq.~\eqref{phase} as 
\begin{equation} \label{partial_cross} \sigma_{\rm abs}^{\ell} = \frac{\pi}{\omega^{2}} (2\ell + 1) \bigl( 1 - |e^{2i\delta_{\omega \ell}}|^{2} \bigr),
\end{equation} 
To calculate the phase shifts, we employ the numerical method described in Refs.~\cite{macedo2016absorption,dolan2009scattering,heidari2024scattering}. We focus on the radial wave equation in Eq.~\eqref{eq:radial2} and compare the numerical solution with the analytical asymptotic expressions~\eqref{Rroman1}--\eqref{Rroman2}. The integration begins at $r\simeq r_{\rm h}+10^{-3}r_{\rm h}$ outside the event range and extends to the asymptotically flat region at $r\simeq 50\, r_{\rm h}$ to obtain a well regulated extraction of the features at large radii. The series used in the asymptotic matching has been truncated at fifteenth order ($N=13$) which is sufficient to achieve numerical convergence at machine precision.
 The effect of regular core scale is studied by computing the absorption cross section for the multipole numbers $\ell=0,1,2$ and different values of \(a/M\) in Fig.~\ref{fig:Psigma}.
At lower frequencies, the monopole part (\(\ell = 0\)) dominates, while higher multipoles contribute as \(\omega\) increases. The standardized regular core scale is within $0$ and $0.5$. There is a correlation between the absorption spectra and the normalized regular core scale parameter. The absorption cross section increases monotonously with the increase of $a/M$ for all the multipole numbers presented. This effect is particularly noticeable at the top of each curve, where the maxima decrease as \(a/M\) increases.

\begin{figure}[ht]
\centering
\includegraphics[width=90mm]{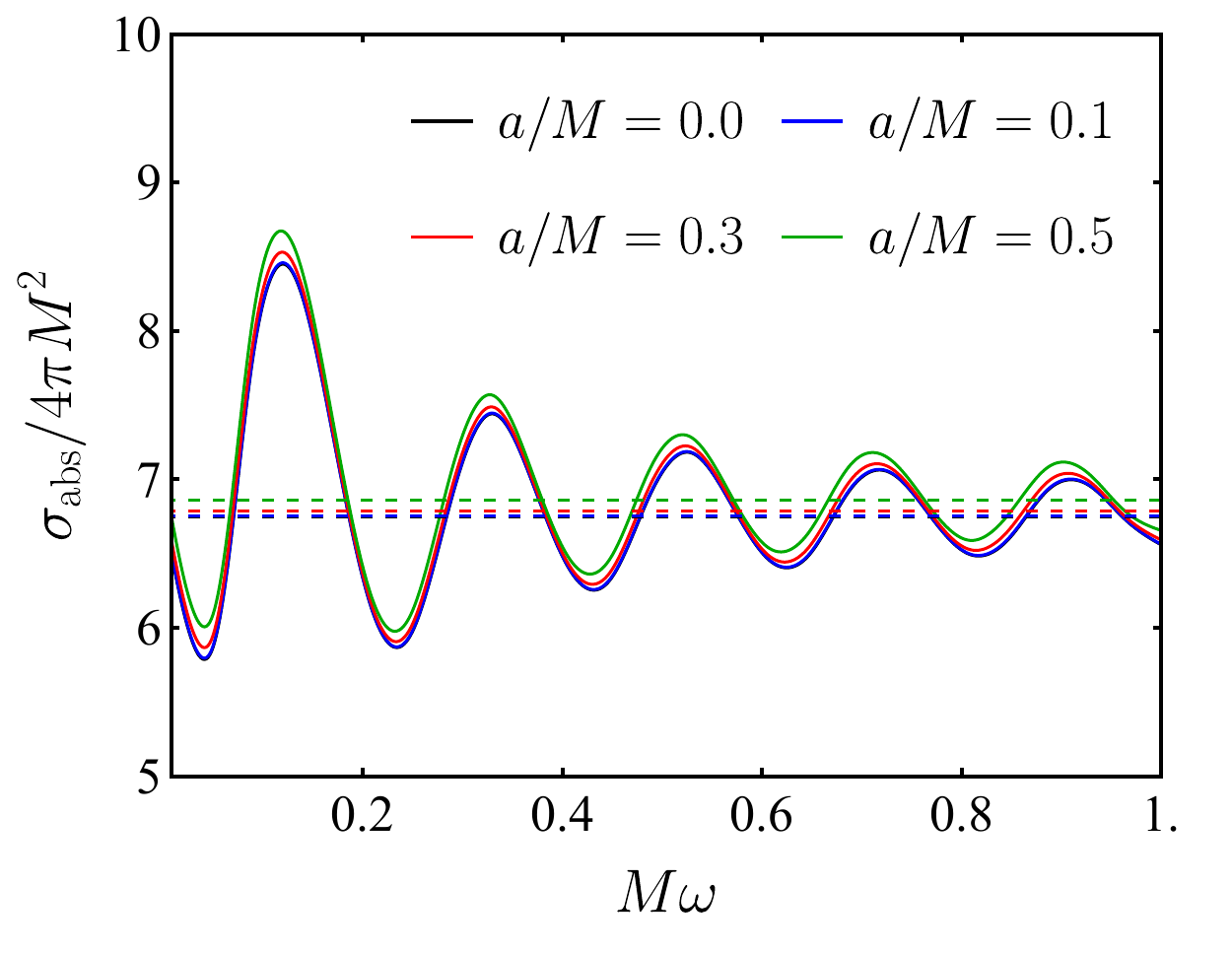}
\caption{Total absorption cross sections as functions of $M\omega$, calculated by summing the partial contributions from $\ell=0$ to $\ell=6$. The dashed lines denote the corresponding geometric cross sections.}
\label{fig:Tsigma}
\end{figure}

Moreover, Fig.~\ref{fig:Tsigma} shows the total absorption cross section obtained by summing the first seven multipoles from $\ell = 0$ to $\ell = 6$. The effect of the normalized regular core scale $a/M$ is clear: we find the total absorption to be enhanced for the whole frequency range with increasing a/M. Note that the overall shape of the curves is basically the same; the position and the locations of the peaks are not affected by the changes in the regular core scale.


Moreover, Fig.~\ref{fig:Tsigma} shows the total absorption cross section, obtained by summing the first seven multipoles from $\ell = 0$ to $\ell = 6$. The influence of the normalized regular core scale $a/M$ is clear: increasing $a/M$ noticeably empowers the total absorption across the entire frequency range. It is worth noting that the overall shape of the curves remains essentially unchanged; neither the position nor the pattern of the peaks is affected by variations of the regular core scale. 


\subsection{Regime of High Frequency{}}

In the short wavelength limit, the absorption process admits a geometric description. Wave propagation is then governed by null geodesics, and the capture cross section is fixed by the critical impact parameter $b_{cr}$ separating photons that escape to infinity from those that cross the event horizon. The geometric absorption cross section is 
\begin{equation}\label{eq:geocross}
\sigma_{\rm geo}=\pi b_{cr}^2.
\end{equation}

In spherical symmetric spacetime, 
the null geodesics can be restricted to the equatorial plane $(\theta=\pi/2,~\dot{\theta}=0)$. The corresponding geodesic Lagrangian is
\begin{equation}
2\mathcal{L}
=-{g_{tt}}\dot{t}^{2}
+{g_{rr}}{\dot{r}^{2}}
+{g_{\phi\phi}}{\dot{\phi}}^{2}
=0,
\end{equation}
where the overdot denotes derivation with respect to an affine parameter. The killing vectors provide two conserved quantities,
\begin{equation}
E=-{g_{tt}}\dot{t},
\qquad
L={g_{\phi\phi}}\dot{\phi},
\end{equation}
representing the photon energy and angular momentum, respectively. Considering the spacetime term of the regular black hole in phanton DBT dark matter, the radial motion can consequently be cast as
\begin{equation}\label{eq:null-radial}
\dot{r}^{,2}+V_{\rm eff}(r)=E^2,
\qquad
V_{\rm eff}(r)=\frac{f(r)L^2}{R(r)^2}.
\end{equation}

The capture threshold is associated with the unstable circular null orbit at $r=r_{\rm ph}$. Applying $V_{\rm eff}=\mathrm{d}V_{\rm eff}/\mathrm{d}r=0$ gives
\begin{equation}\label{cr}
2R'(r_{\rm ph})f(r_{\rm ph})-R(r_{\rm ph})f'(r_{\rm ph})=0.
\end{equation}
The critical impact parameter then follows as
\begin{equation}\label{impact}
b_{cr}
=\frac{L}{E}
=\frac{R(r_{\rm ph})}{\sqrt{f(r_{\rm ph})}}.
\end{equation}

Substituting the exact metric function in Eq.~\eqref{cr}, the unstable photonic radius will be 
$r_{\mathrm{ph}} = 3M$. Thus, the impact parameter obtained from Eq. \eqref{impact} takes the approximated compact form
\begin{equation}\label{sigmahigh}
b_{cr}
\simeq
3\sqrt{3} M
\left[
1+\frac{1}{30}\left(\frac{a}{M}\right)^2
+\mathcal{O}\left({\frac{a}{M}}\right)^4
\right].
\end{equation}
Substituting the result in Eq.~\eqref{eq:geocross}, the geometric cross section, in the perturbative regime, this expression becomes
\begin{equation}\label{eq:geo-expansion3}
\sigma_{\rm geo}
\simeq
27\pi M^2
\left[
1+\frac{1}{15}\left(\frac{a}{M}\right)^2
+\mathcal{O}\left({\frac{a}{M}}\right)^4
\right].
\end{equation}

Therefore, the Schwarzschild capture cross section $27\pi M^2$ is recovered for $a/M\rightarrow0$, while a finite regular core scale produces a positive correction.

Fig.~\ref{fig:Tsigma} compares the partial wave results with the corresponding geometric absorption cross sections. The numerical total cross section is obtained by truncating the multipole sum at $\ell_{\max}=6$, and the values calculated from Eq.~\eqref{sigmahigh} are shown as dashed horizontal lines. At high frequencies, the numerical curves oscillate around their respective geometric capture limits. As pointed out, increasing $a/M$ raises both the geometric optics value and the envelope of the numerical cross section with convergence to the geometric cross section.

\subsection{Regime of Low Frequencies{}}

At sufficiently low frequencies, the absorption cross section of a black hole is assumed to approach the area of its event horizon. The literature has extensively established this universal behavior, which is a robust characteristic of black hole scattering processes \cite{Das1996we,heidari2025absorption,Higuchi:2001si}. In the long-wavelength regime, the detailed structure of the spacetimshowss subdominant, as the incident wave probes length scales that are comparable to or larger than the size of a black hole. Consequently, the absorption mechanism is predominantly dictated by the geometric characteristics of the horizon, leading the cross section to asymptotically converge with the horizon area.

Specifically, for scalar field fluctuations propagating on a Schwarzschild background, it is analytically demonstrated that the low frequency absorption cross section aligns with the geometric region of the event horizon. Notably, this constraint remains applicable across various situations, such as rotating black holes, modified black hole characteristics, and complex surrounding geometries, as long as the incoming frequency is adequately low \cite{leite2017scalar,baptista2025scattering,macedo2013absorption,heidari2024scattering,heidari2026particle}.

Consequently, in the low--frequency regime, the absorption cross section can be expressed in terms of the radius of the event horizonas $r_{\text{h}}$ as 
\begin{align}\label{eq:lowfr} 
\sigma_{\text{low}} & \simeq 4\pi R(r_{\text{h}})=4\pi(r_{\text{h}}^2+a^2), 
\end{align} 
where the black hole radius is obtained by imposing $f(r_\text{h})=0$ and has the following form as \begin{equation} 
r_\text{h}=\frac{M}{3} \Big(1+\mathcal{A}+\frac{1}{\mathcal{A}}\Big) 
\end{equation} $\mathcal{A}$ is defined as $\Big[1-\frac{27}{40} (\frac{a}{M})^2+\frac{3}{40} (\frac{a}{M}) \sqrt{ 81 (\frac{a}{M})^2-240}\Big]^{1/3}$. 
If $\sigma_{\text{low}}$ is expanded in terms of the small parameter $a/M$ up to the second order, we arrive at the following expression 
\begin{align}\label{eq:lowfr} 
\sigma_{\text{low}} & \simeq 16 \pi M^2\Big[1+\frac{12}{5} (\frac{a}{M})^2\Big]. 
\end{align} 

\begin{figure}[ht]
    \centering
    \includegraphics[width=90mm]{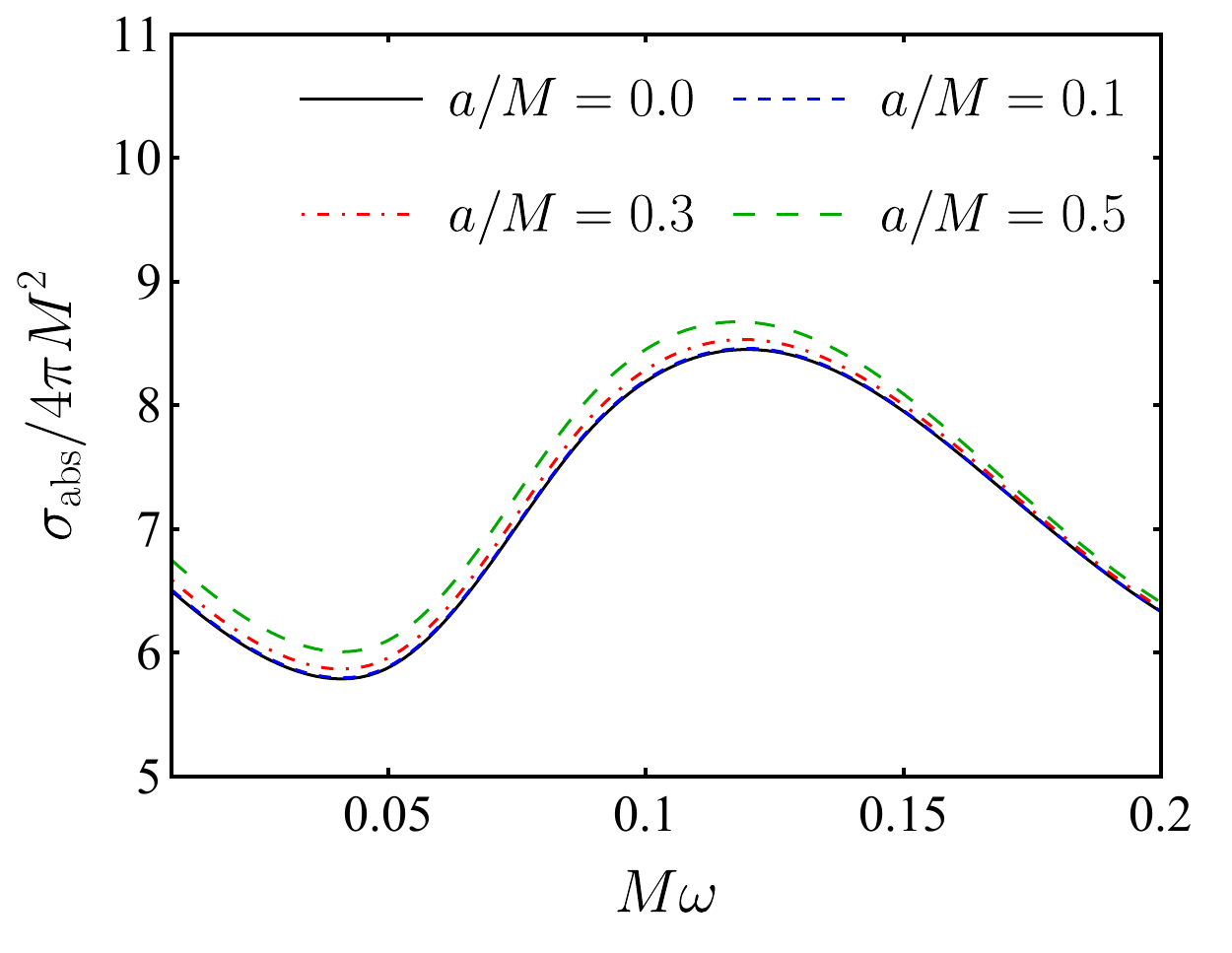}
    \caption{Low frequency part of the total absorption cross sections for various values of the regular core scale parameter.}
    \label{fig:TsigmaLow}
\end{figure}

The low--frequency behavior of the absorption cross section for different values of the normalized regular core scale is illustrated in Fig.~\ref{fig:TsigmaLow}. The absorption cross section exhibits sensitivity to variations in $a/M$. Increasing $a/M$ leads to an increase of the absorption cross section, which is consistent with the behavior expected in Eq.~\ref{eq:lowfr}.

\section{Scattering cross section{}}

The scattering of waves by a black hole can be systematically described using a partial wave expansion
\cite{futterman1986scattering,dolan2013scattering,dolan2009scattering,anacleto2020absorption,leite2019black,baptista2025scattering}.
Within this framework, the scattering amplitude is expressed as
\begin{equation}
g(\theta) =
\frac{1}{2i\omega}
\sum_{\ell=0}^{\infty}
(2\ell+1)\left( e^{2i\delta_{\ell}} - 1 \right)
P_{\ell}(\cos\theta),
\label{eq:amplitude}
\end{equation}
where $P_\ell(\cos\theta)$ denotes the Legendre polynomial of degree $\ell$, and $\delta_\ell$ is the phase shift associated with the $\ell$th partial wave, as defined in Eq.~\eqref{phase}. The entire angular dependence of the scattering process is encoded in the phase shifts through this expansion.

The observable angular distribution of the scattered radiation is characterized by the differential scattering cross section, which is related to the scattering amplitude by \cite{sanchez1978elastic}
\begin{equation}
\frac{\mathrm{d}\sigma}{\mathrm{d}\Omega} = \left| g(\theta) \right|^2.
\label{eq:diff_cross_section}
\end{equation}
This quantity provides direct information about the angular structure of the scattering process.

The total scattering cross section is obtained by integrating the differential cross section over the full solid angle. For a monochromatic wave of frequency $\omega$, this leads to
\begin{equation}
\sigma_{\text{sca}} =
\int \frac{\mathrm{d}\sigma}{\mathrm{d}\Omega}\, \mathrm{d}\Omega
= \frac{\pi}{\omega^2}
\sum_{\ell=0}^{\infty}
(2\ell+1)
\left| e^{2i\delta_\ell} - 1 \right|^2.
\label{eq:total_cross_section}
\end{equation}
This expression highlights that the total scattering cross section arises from a coherent superposition of all partial waves, with each contribution weighted by the corresponding phase shift.  


\subsection{Partial Scattering cross section{}} 
The partial wave decomposition is a useful technique to study the contribution of each angular momentum separately. The mode analysis allows us to estimate the sensitivity of each multipole to the regular core score parameter. Based on Eq.~\eqref{eq:amplitude}, the corresponding partial differential scattering cross section is defined as \begin{equation}\label{eq:partial_cross_section} \frac{\mathrm{d}\sigma_{\rm sca}^{\ell}}{\mathrm{d}\Omega}= \left| \frac{2\ell+1}{2i\omega} \left(e^{2i\delta_\ell}-1\right) P_\ell(\cos\theta) \right|^2. 
\end{equation} 
For fixed $\ell$ and $\omega$, the phase shift $\delta_\ell$ is the term influenced by the background geometry. Thus, the variation of the dimensionless regular core score parameter $a/M$ alters the magnitude of the partial cross section, but its nodal structure is still given by the Legendre polynomial $P_\ell(\cos\theta)$.

\begin{figure}[ht!]
\centering
\includegraphics[width=90mm]{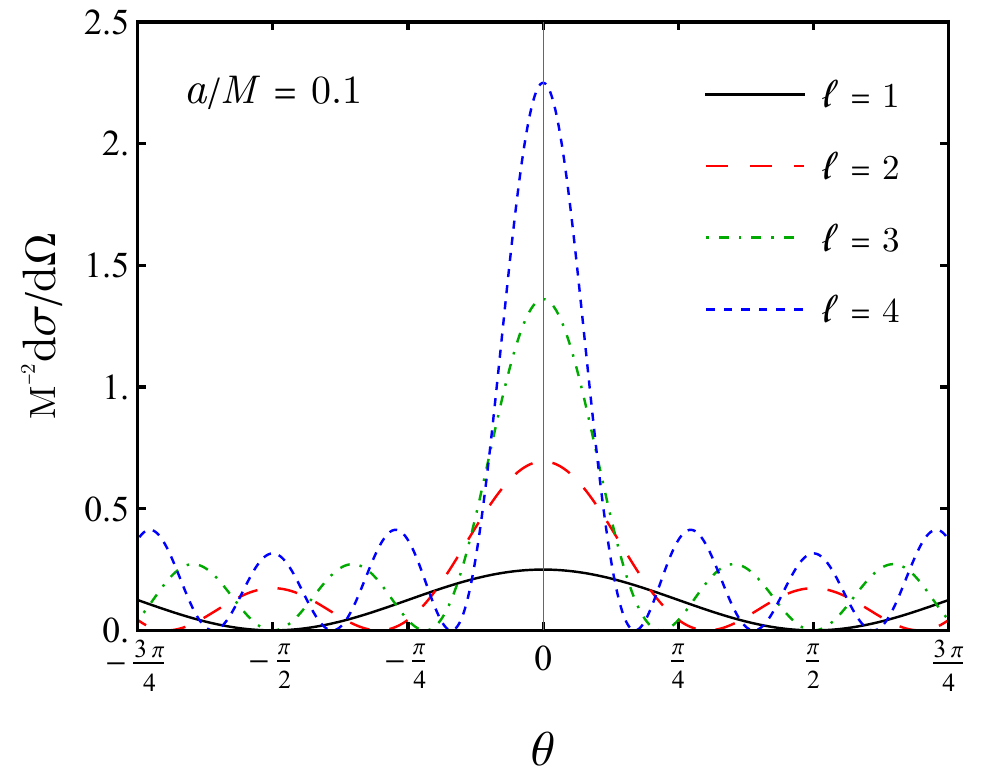}
\caption{Normalized partial differential scattering cross sections for $\ell=1,\ldots,4$, and the fixed regular core score $a/M=0.1$.}
\label{fig:Sct01}
\end{figure}
\begin{figure}[ht!] \centering \includegraphics[width=85mm]{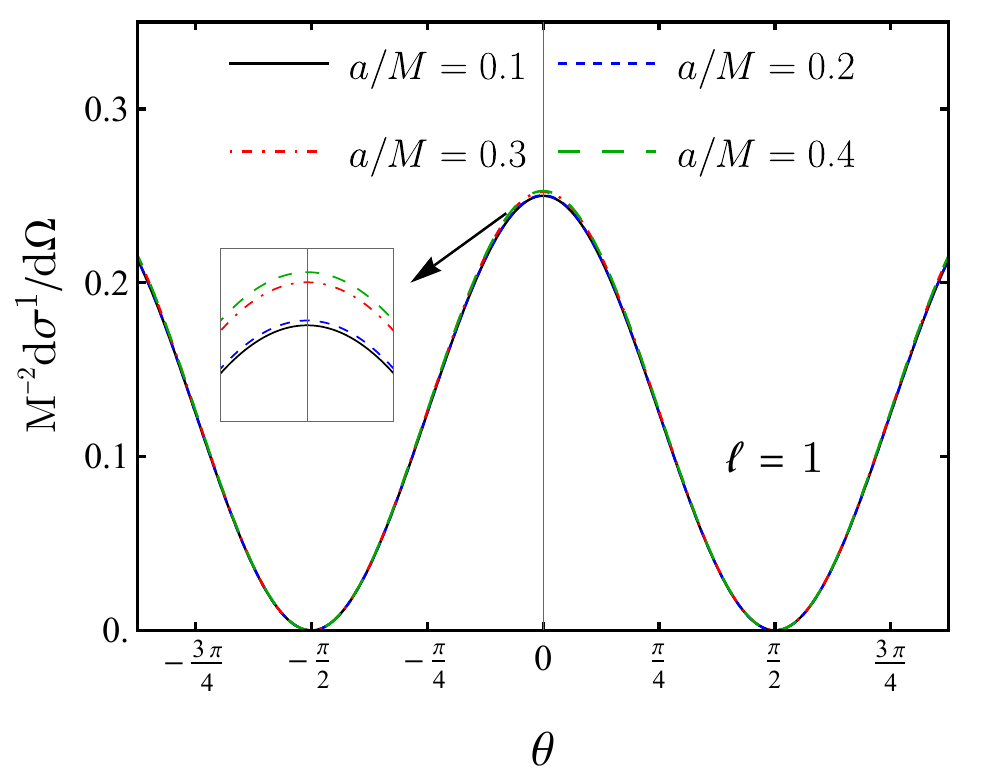} \includegraphics[width=85mm]{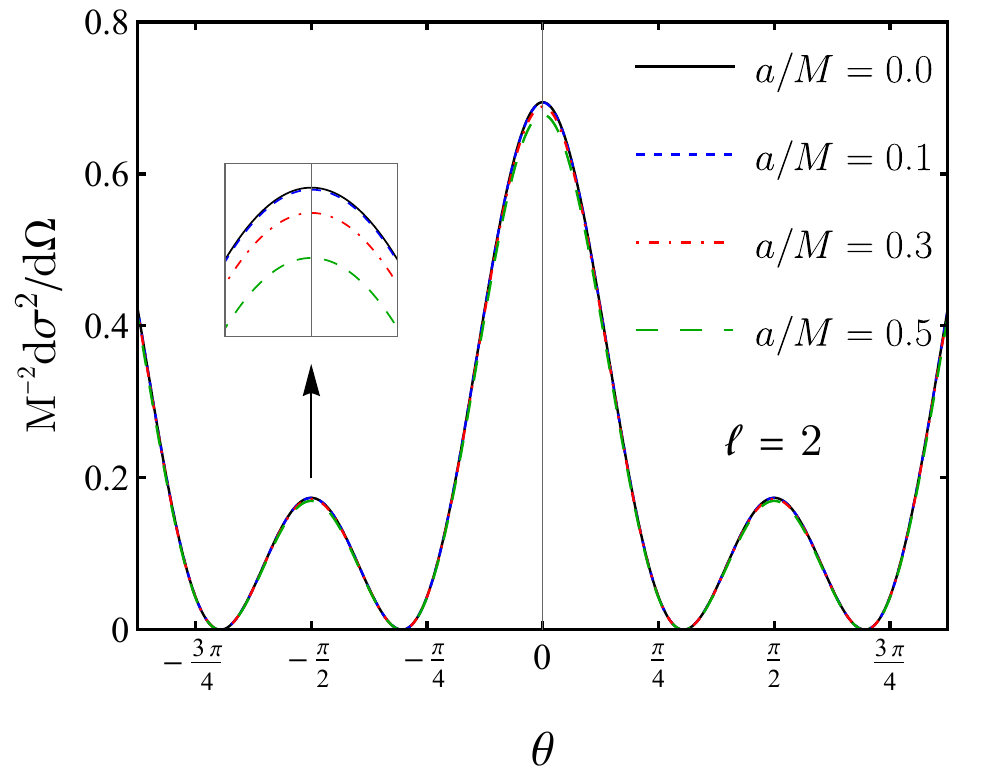} \includegraphics[width=85mm]{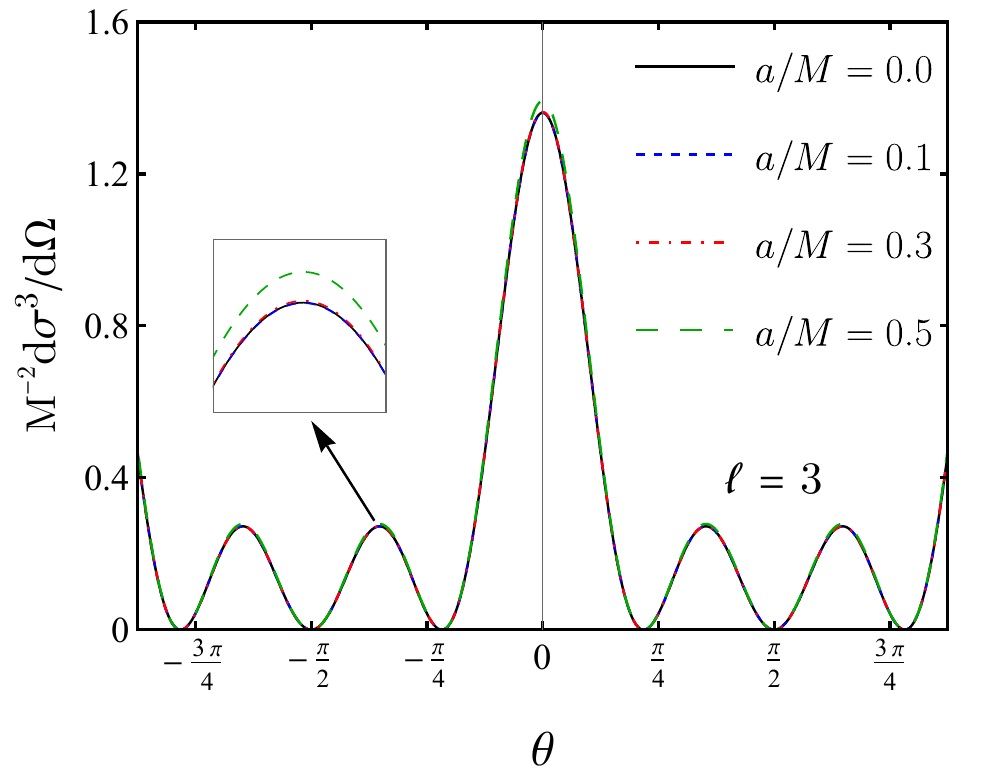} \caption{Normalized partial differential scattering cross sections at $M\omega=3$ for several values of $a/M$. The panels are ordered from top to bottom from $\ell=1$ to $\ell=3$.} \label{fig:Sct02} \end{figure}

Fig.~\ref{fig:Sct01} compares the first four nonzero multipoles at fixed $a/M=0.1$. We examine the regular core scale dependence at fixed frequency $M\omega=3$ in Fig.~\ref{fig:Sct02}. The three panels are for $\ell=1$, $2$ and $3$, respectively. For a fixed multipole, the change of $a/M$ does not affect the positions of the angular peaks but affects the amplitude through the associated phase shift. The first two odd modes exhibit a mild increase of the amplitude with the increase of $a/M$, while the behavior of the $\ell=2$ contribution is the opposite. The response of the multipole dependence indicates that the regular core score parameter does not affect all the phase shifts in the same way.

Because the different multipoles scale differently with $a/M$, the behavior of the total differential cross section cannot be deduced from a simple comparison of their individual magnitudes. The total angular profile is the coherent sum in Eq. \eqref{eq:amplitude} and thus also contains interference terms between different multipoles, which will be discussed in the next section.

\subsection{Total scattering cross section{}}
Numerical evaluation of the scattering amplitude via the partial wave expansion in Eq. \eqref{eq:amplitude} exhibits poor convergence due to the oscillatory nature of the Legendre polynomials. To obtain accurate numerical results, we employ a convergence acceleration method, originally developed by Yennie et al. \cite{yennie1954phase} and later adapted to black hole scattering contexts \cite{dolan2006fermion}. This iterative regularization procedure systematically redefines the series coefficients, effectively suppressing high $\ell$ oscillations and yielding a rapidly convergent representation. We used second--order regularized series; convergence is obtained for the studied frequency range by including partial waves up to $\ell \thicksim 50$.
\begin{figure}[ht!]
	\centering
	\includegraphics[width=85mm]{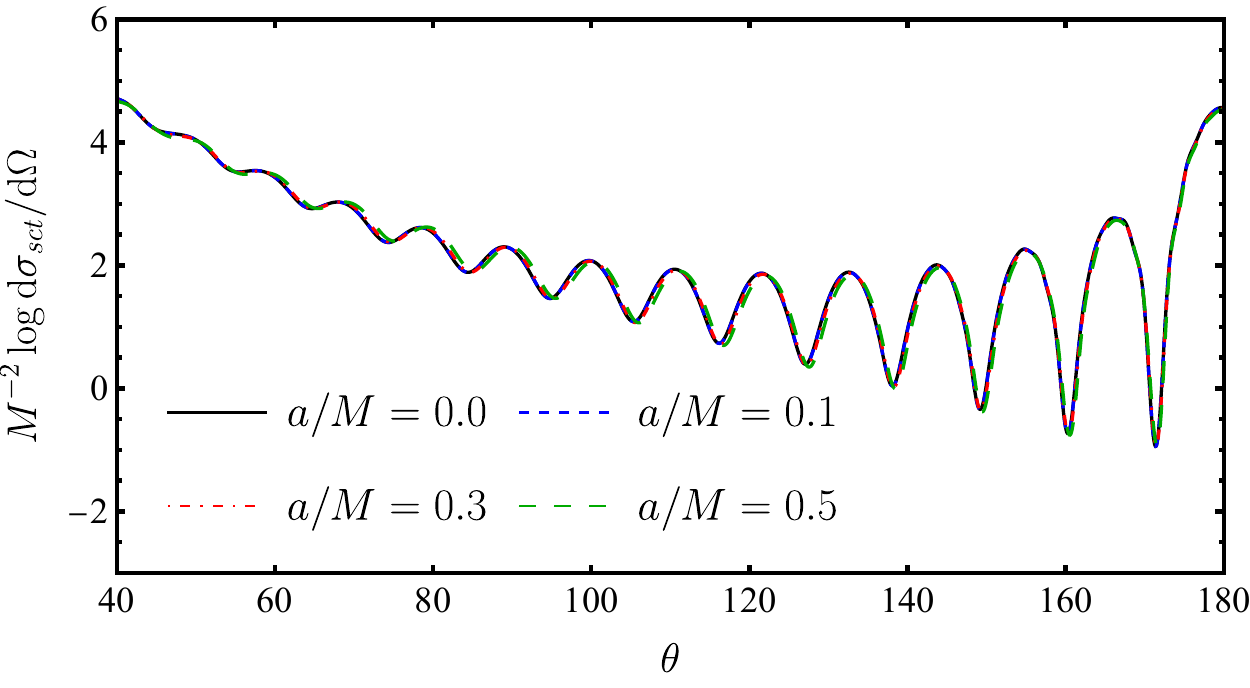}   \includegraphics[width=85mm]{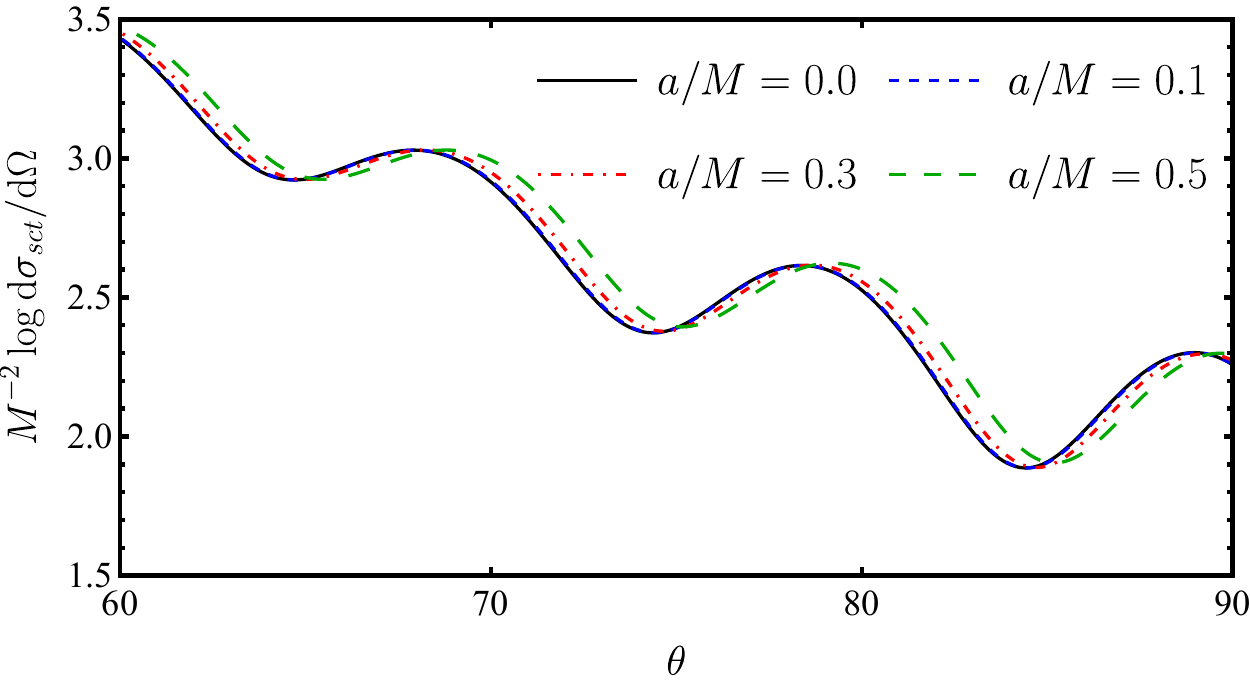} 
    \includegraphics[width=85mm]{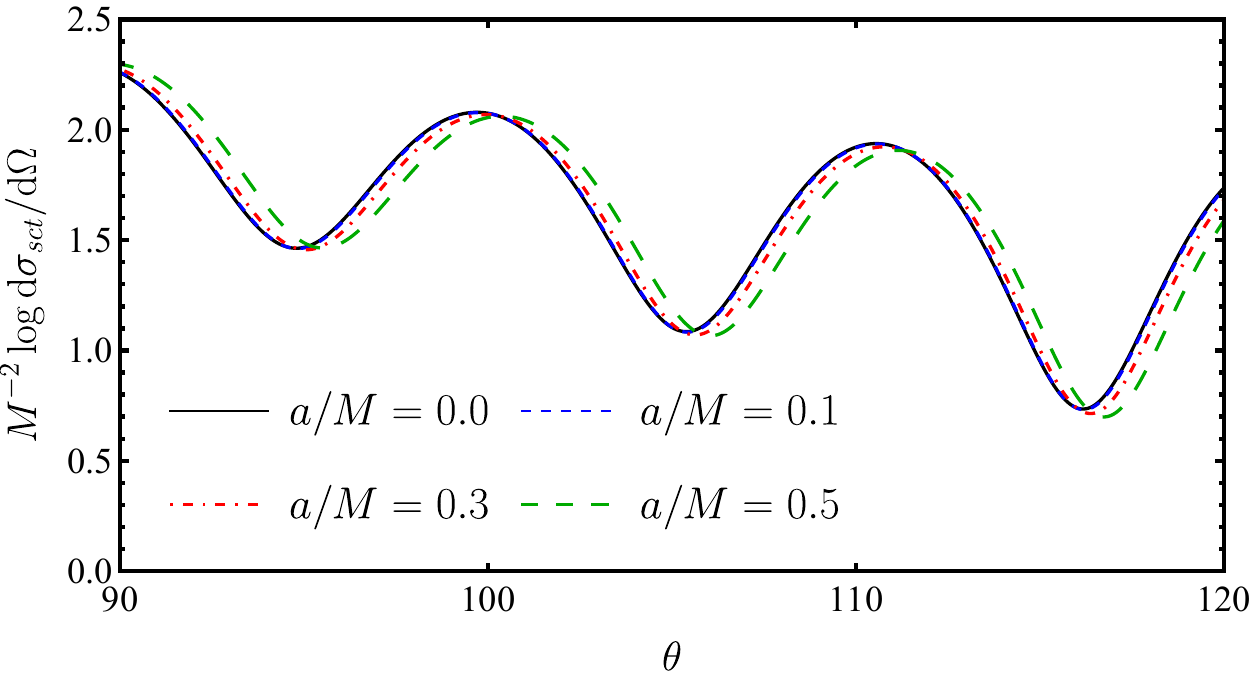} 
    \includegraphics[width=85mm]{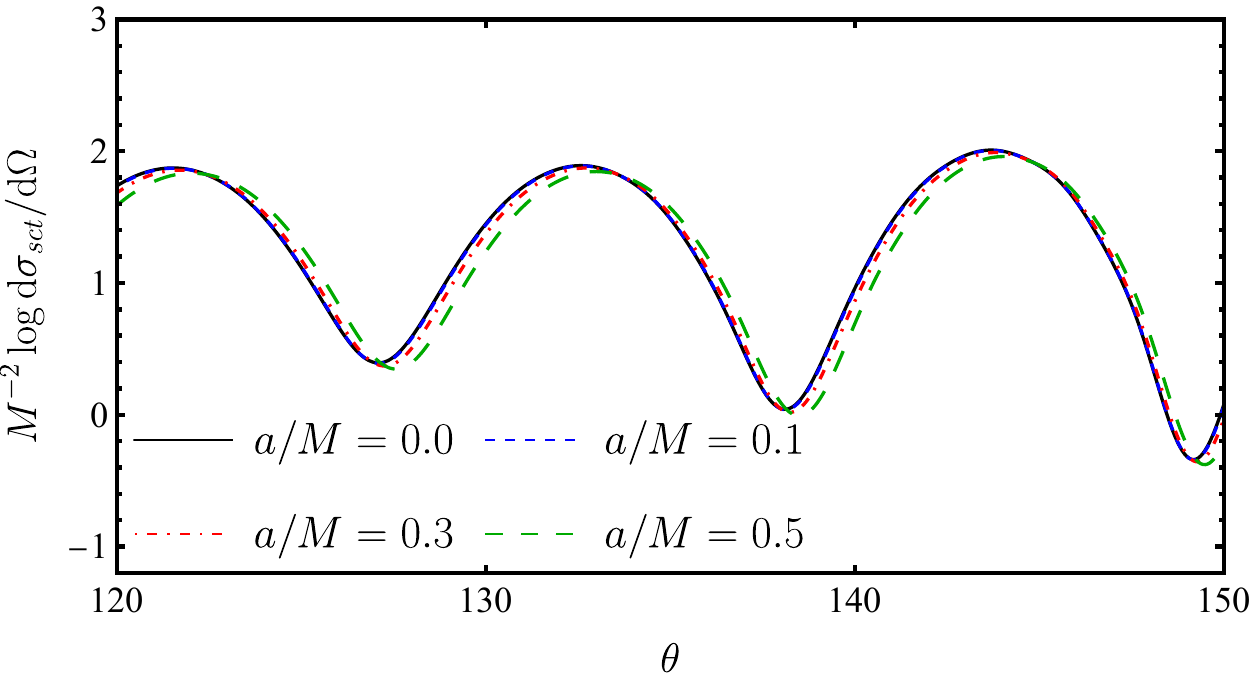}
	\caption{ Total scattering cross section in logarithmic scale at fixed frequency $M\omega = 3$.
Upper panel: Dependence on the density parameter $a/M$, varied from $0$ to $0.5$.
Middle and Lower panels: Focused plots for better visualization in a zoomed domain.}
	\label{fig:Tscat}
\end{figure}
The angular dependence of the total scattering cross section, displayed on a logarithmic scale as a function of the scattering angle $\theta$, is shown in Fig.~\ref{fig:Tscat}. The middle and lower panels are zoomed versions of the domain $60-90$, $90-120$, and $120-150$ degree, and reveals that increasing the scaled regular core parameter $a/M$ results in different amplitude behaviour in different intervals. In the lower domain $60-90$ degrees, the amplitude of total scattering has a subtle increase; on the other hand, in the domain $120-150$, the amplitude decreases. However, there is a general trend in the shift of the interfering fringe patterns toward the higher degrees.


\section{Geodesics}

Geodesic trajectories provide a direct probe of the strong field geometry, independently of the wave analysis developed in the preceding sections. In the present spacetime, the parameter $a$ modifies the lapse function and hence the connection that governs freely falling particles. Comparing null and timelike trajectories therefore offers a simple way to assess how the regular core deformation influences photon propagation and the motion of massive test bodies.

For an affinely parametrized trajectory $x^\mu(\lambda)$, the equations of motion are
\begin{equation}\label{Eq:Geo}
\frac{\mathrm{d}^{2}x^\mu}{\mathrm{d}\lambda^{2}}
+\Gamma^\mu_{\alpha\beta}
\frac{\mathrm{d}x^\alpha}{\mathrm{d}\lambda}
\frac{\mathrm{d}x^\beta}{\mathrm{d}\lambda}
=0,
\end{equation}
where $\Gamma^\mu_{\alpha\beta}$ are the Christoffel symbols associated with the metric. The distinction between null and timelike motion is imposed through the normalization condition
\begin{equation}\label{eq:geodesicNormalization}
g_{\mu\nu}
\frac{\mathrm{d}x^\mu}{\mathrm{d}\lambda}
\frac{\mathrm{d}x^\nu}{\mathrm{d}\lambda}
=\varepsilon,
\qquad
\varepsilon=
\begin{cases}
0: & \text{lightlike geodesics},\\ -1: & \text{timelike geodesics}.
\end{cases}
\end{equation}

For the static and spherically symmetric line element considered here,
the four component equations following from Eq.~\eqref{Eq:Geo} can be expressed as
\begin{align}\label{geot}
&\frac{\mathrm{d}^2 t}{\mathrm{d}{\lambda}^2} +
\frac{2 M \left(5 r^2-3 a^2\right) r' t'}{2 a^2 M r+5 r^3 (r-2 M)}=0,\\ \label{geor}
&\frac{\mathrm{d}^2r}{\mathrm{d}{\lambda}^2} -\frac{M \left(5 r^2-3 a^2\right) r'^2}{2 a^2 M r+5 r^3 (r-2 M)}\\  \nonumber
&-\frac{\left(2 a^2 M+5 r^2 (r-2 M)\right) \left(M \left(3 a^2-5 r^2\right) t'^2+5 r^5 \theta '^2+\sin ^2\theta \varphi '^2\right)}{25 r^7}=0,\\ \label{geotheta}
&\frac{\mathrm{d}\theta^{\prime}}{\mathrm{d}{\lambda}}-\sin \theta  \cos \theta  \varphi '^2+\frac{2 \theta ' r'r}{r^2+a^2}=0,\\ \label{geophi}
&\frac{\mathrm{d}\varphi^{\prime}}{\mathrm{d}{\lambda}}+\frac{2 \varphi ' r'r}{r^2+a^2}+2 \varphi ' \theta ' \cot \theta  =0.
\end{align}
where an overdot denotes differentiation with respect to $\lambda$. Spherical symmetry allows the motion to be restricted to the equatorial plane by choosing $\theta=\pi/2$ and $\dot{\theta}=0$ initially. The remaining equations are integrated simultaneously, with initial data satisfying Eq.~\eqref{eq:geodesicNormalization}.

\begin{figure*}[t]
\centering
\includegraphics[width=0.48\textwidth]{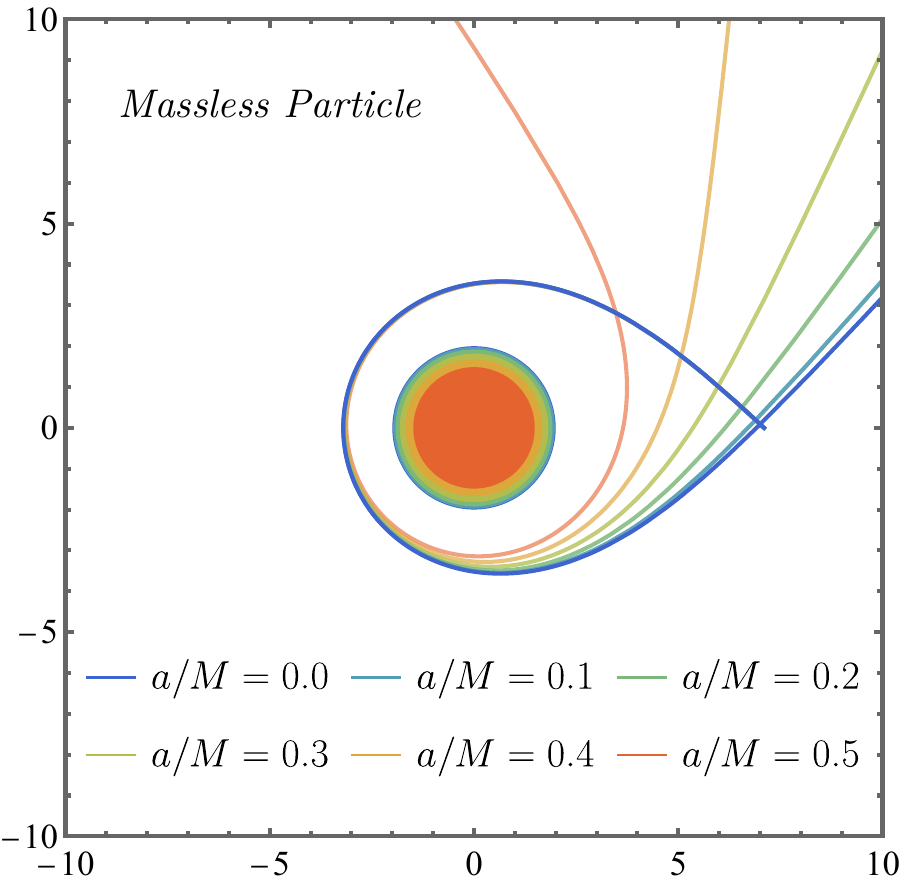}
\includegraphics[width=0.48\textwidth]{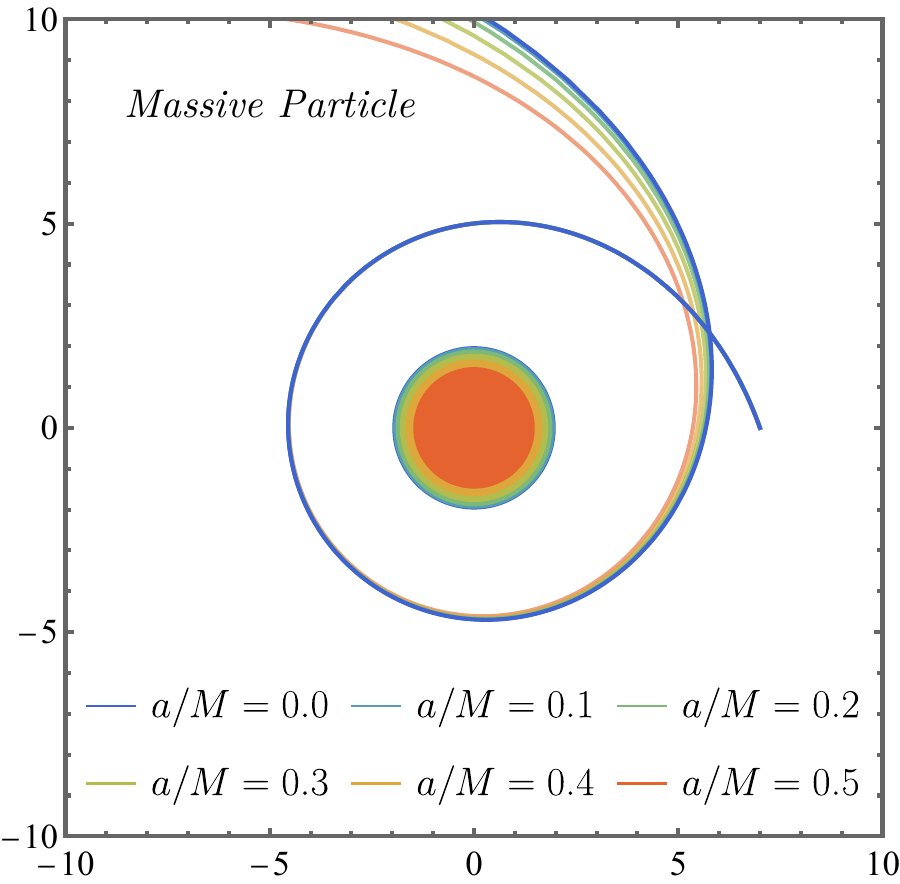}
\caption{Lightlike (left) and timelike geodesics (right) for selected values of $a/M$ in the interval $a/M=0.0$ to $0.5$. The initial conditions are held fixed in each comparison.}
\label{fig:Geo}
\end{figure*}

Fig.~\ref{fig:Geo} compares null and timelike geodesics for selected values of the dimensionless core size parameter $a/M$ with different colors. Furthermore, the colored circle represents the black hole, with colors corresponding to each $a/M$ value. The same initial conditions are used within each panel so that differences between the trajectories can be attributed directly to the regular deformation. As discussed in Ref. \cite{parvez2026exact}, increasing the parameter $a/M$ leads to a slight decrease in the event horizon radius. The numerical trajectories show a systematic increase in the deflection as $a/M$ increases. For the initial conditions adopted here, both massless and massive particles remain closer to the central object when the regular core contribution is enhanced.

\section{Conclusion}

The present study provided a unified characterization of particle emission, evaporation, scalar wave propagation, and geodesic motion in the regular black hole supported by a phantom Dirac--Born--Infeld field. The analysis combined quantum field theoretic, semiclassical, partial wave, and geodesic methods. Throughout the calculations, the ratio $a/M$ controlled the regular deformation, whereas the limit $a\rightarrow0$ recovered the Schwarzschild results, as we should expect.

The fixed background Bogoliubov treatment showed that scalar and massless fermionic fields followed Bose--Einstein and Fermi--Dirac distributions, respectively, while both sectors were governed by the same surface gravity. The DBI correction lowered the Hawking temperature and reduced the occupation of every mode with nonzero frequency. The agreement between the bosonic and fermionic temperatures confirmed the geometric origin of the thermal factor, whereas the distinct statistical denominators followed from the corresponding commutation and anticommutation relations.

When energy conservation was imposed through the tunneling prescription, the emission spectrum ceased to be exactly thermal. The replacement of the black hole mass by its instantaneous value along the tunneling trajectory generated the usual recoil term and an additional logarithmic correction associated with the DBI scale. The low--energy expansion reproduced the temperature obtained from the Bogoliubov coefficients, while the higher--order terms retained the correlations induced by the changing mass. At fixed $M$ and $\omega$, increasing $a$ raised the tunneling exponent and suppressed both bosonic and fermionic emission.

The evaporation rate was formulated as a sum over particle species and angular channels, with the corresponding greybody factors retained. In the geometric--optics regime, the coordinate position of the unstable photon orbit remained $r_{\mathrm{ph}}=3M$, although the critical impact parameter and the cross section increased with the regular core scale. Once this enlargement was combined with the modified temperature, the thermal suppression prevailed: the luminosity decreased and the evaporation time acquired a positive correction at order $\mathcal{O}(a^{2})$. This result remained restricted to the controlled domain $M\gg a$. The exact parametrization of the geometry showed that the lower mass boundary $M=2a/(3\pi)$ did not coincide with a vanishing Hawking temperature. The perturbative mass loss law could therefore not be extrapolated to that boundary, and the existence of a stable remnant was not established by the semiclassical evaporation analysis alone.

For the wave sector, the massless Klein--Gordon equation was reduced to a one--dimensional scattering problem whose effective potential vanished at the horizon and at spatial infinity. Numerical matching provided the reflection and transmission amplitudes, from which the phase shifts and cross sections were obtained. The monopole governed the low--frequency absorption, while progressively higher multipoles contributed as the frequency increased. The total absorption cross section grew with $a/M$ throughout the investigated range. In the long-wavelength limit, it approached the horizon area $4\pi R^{2}(r_h)$; at high frequencies, it oscillated around the enlarged geometric capture cross section. The principal spectral structure remained nearly unchanged, although its magnitude was enhanced by the regular deformation.

The scattering phase shifts displayed a genuinely multipolar response. At $M\omega=3$, the $\ell=1$ and $\ell=3$ contributions developed a mild enhancement as $a/M$ increased, whereas the $\ell=2$ amplitude was reduced. Their coherent superposition redistributed the scattered intensity across the angular profile: the cross section increased slightly between $60^\circ$ and $90^\circ$, decreased between $120^\circ$ and $150^\circ$, and exhibited interference fringes displaced toward larger angles. The angular nodes of each isolated multipole remained fixed by the corresponding Legendre polynomial.

The numerical integration of the geodesic equations completed the strong field analysis. Under identical initial conditions, increasing $a/M$ slightly reduced the coordinate radius of the event horizon and enhanced the deflection of both null and timelike trajectories. The particles consequently remained closer to the central object when the regular core contribution became stronger. The regular deformation therefore acted differently on horizon emission and exterior propagation: it weakened the radiative loss, enlarged the absorption cross section, reshaped the scattering pattern, and strengthened the bending of massless and massive particles.

As a natural extension, it would be worthwhile to investigate, for this black hole considered in this paper, gravitational lensing in both the weak-- and strong--deflection limits, along the lines of Refs.~\cite{AraujoFilho:2025hnf,AraujoFilho:2024mvz,Filho:2024tgy,Filho:2024isd}, as well as from the perspective of ensemble theory \cite{Filho:2021zuf,furtado2023thermal,AraujoFilho:2026tvy}.

\section*{Acknowledgments}
\hspace{0.5cm}

N. H. is supported by Conselho Nacional de Desenvolvimento Cient\'{\i}fico e Tecnol\'{o}gico — CNPq, project number 152891/2025-0. Also, N. H. is grateful for the support provided by three COST Actions: CA21106 (COSMIC WISPers in the Dark Universe: Theory, Astrophysics and Experiments), CA21136 (Addressing Observational Tensions in Cosmology with Systematics and Fundamental Physics, also known as CosmoVerse), and CA23130 (Bridging High and Low Energies in Search of Quantum Gravity, or BridgeQG). In addition, A. A. Araújo Filho is supported by Conselho Nacional de Desenvolvimento Cient\'{\i}fico e Tecnol\'{o}gico (CNPq) with project number 150223/2025-0. V. B. B. is partially supported by the National Council for Scientific and Technological Development ( CNPq)- Brazil, Grant number 311847/2026-9. I. P. L. acknowledges partial support from the National Council for Scientific and Technological Development, CNPq, under grant 312547/2023-4. I. P. L. acknowledges the networking support by the COST Action BridgeQG (CA23130), the COST Action RQI (CA23115) and the COST Action FuSe (CA24101) supported by COST (European Cooperation in Science and Technology). Additionally, we express our gratitude to Prof. Caio F. B. Macedo and Prof. S. Dolan for their insightful discussions regarding the scattering cross section.


	\bibliography{main}

\begin{thebibliography}{10}

\bibitem{penrose1965gravitational}
Roger Penrose.
\newblock Gravitational collapse and space-time singularities.
\newblock {\em Physical Review Letters}, 14(3):57, 1965.

\bibitem{penrose1978singularities}
Roger Penrose.
\newblock Singularities of spacetime.
\newblock {\em Theoretical principles in astrophysics and relativity}, pages
  217--243, 1978.

\bibitem{hawking1970singularities}
Stephen~William Hawking and Roger Penrose.
\newblock The singularities of gravitational collapse and cosmology.
\newblock {\em Proceedings of the Royal Society of London. A. Mathematical and
  Physical Sciences}, 314(1519):529--548, 1970.

\bibitem{Filho:2023qxu}
A.~A.~Ara{\'u}jo Filho.
\newblock {Implications of a Simpson{\textendash}Visser solution in
  Verlinde{\textquoteright}s framework}.
\newblock {\em Eur. Phys. J. C}, 84(1):73, 2024.

\bibitem{Filho:2023voz}
A.~A.~Ara{\'u}jo Filho.
\newblock {Analysis of a regular black hole in Verlinde{\textquoteright}s
  gravity}.
\newblock {\em Class. Quant. Grav.}, 41(1):015003, 2024.

\bibitem{Bardeen1968}
James~M. Bardeen.
\newblock Non-singular general-relativistic gravitational collapse.
\newblock In {\em Proceedings of the International Conference GR5}, page 174,
  Tbilisi, USSR, 1968.

\bibitem{AyonBeatoGarcia1998}
Eloy Ay{\'o}n-Beato and Alberto Garc{\'i}a.
\newblock Regular black hole in general relativity coupled to nonlinear
  electrodynamics.
\newblock {\em Phys. Rev. Lett.}, 80:5056--5059, 1998.

\bibitem{AyonBeatoGarcia1999}
Eloy Ay{\'o}n-Beato and Alberto Garc{\'i}a.
\newblock New regular black hole solution from nonlinear electrodynamics.
\newblock {\em Phys. Lett. B}, 464:25--29, 1999.

\bibitem{AyonBeatoGarcia2005}
Eloy Ay{\'o}n-Beato and Alberto Garc{\'i}a.
\newblock Four-parametric regular black hole solution.
\newblock {\em Gen. Relativ. Gravit.}, 37:635--641, 2005.

\bibitem{AraujoFilho:2026hun}
A.~A. Ara{\'u}jo~Filho, Ednaldo~L. B., Junior., Jos{\'e} Tarciso~S. S.,
  Junior., Francisco S.~N. Lobo, Jorde A.~A. Ramos, Manuel~E. Rodrigues, Diego
  Rubiera-Garcia, Lu{\'\i}s F.~Dias da~Silva, and Henrique~A. Vieira.
\newblock {Regular black holes in general relativity from nonlinear
  electrodynamics with de Sitter cores}.
\newblock {\em Phys. Dark Univ.}, 53:102396, 2026.

\bibitem{Bronnikov2001}
Kirill~A. Bronnikov.
\newblock Regular magnetic black holes and monopoles from nonlinear
  electrodynamics.
\newblock {\em Phys. Rev. D}, 63:044005, 2001.

\bibitem{BalartVagenas2014}
Leonardo Balart and Elias~C. Vagenas.
\newblock Regular black holes with a nonlinear electrodynamics source.
\newblock {\em Phys. Rev. D}, 90:124045, 2014.

\bibitem{BurinskiiHildebrandt2002}
Alexander Burinskii and Sergi~R. Hildebrandt.
\newblock New type of regular black holes and particlelike solutions from
  nonlinear electrodynamics.
\newblock {\em Phys. Rev. D}, 65:104017, 2002.

\bibitem{Culetu2015Nonsingular}
Hristu Culetu.
\newblock Nonsingular black hole with a nonlinear electric source.
\newblock {\em Int. J. Mod. Phys. D}, 24:1542001, 2015.

\bibitem{FanWang2016}
Zhong-Ying Fan and Xiaobao Wang.
\newblock Construction of regular black holes in general relativity.
\newblock {\em Phys. Rev. D}, 94:124027, 2016.

\bibitem{shankaranarayanan2004non}
S~Shankaranarayanan and Naresh Dadhich.
\newblock Non-singular black-holes on the brane.
\newblock {\em International Journal of Modern Physics D}, 13(06):1095--1103,
  2004.

\bibitem{HuLanMiao2023}
Han-Wen Hu, Chen Lan, and Yan-Gang Miao.
\newblock A regular black hole as the final state of evolution of a singular
  black hole.
\newblock {\em Eur. Phys. J. C}, 83:1047, 2023.

\bibitem{BronnikovFabris2006}
Kirill~A. Bronnikov and J{\'u}lio~C. Fabris.
\newblock Regular phantom black holes.
\newblock {\em Phys. Rev. Lett.}, 96:251101, 2006.

\bibitem{Hayward2006}
Sean~A. Hayward.
\newblock Formation and evaporation of nonsingular black holes.
\newblock {\em Phys. Rev. Lett.}, 96:031103, 2006.

\bibitem{Dymnikova1992}
Irina Dymnikova.
\newblock Vacuum nonsingular black hole.
\newblock {\em Gen. Relativ. Gravit.}, 24:235--242, 1992.

\bibitem{uchikata2012new}
Nami Uchikata, Shijun Yoshida, and Toshifumi Futamase.
\newblock New solutions of charged regular black holes and their stability.
\newblock {\em Physical Review D—Particles, Fields, Gravitation, and
  Cosmology}, 86(8):084025, 2012.

\bibitem{VertogradovOvgun2025}
Vitalii Vertogradov and Ali {\"O}vg{\"u}n.
\newblock Exact regular black hole solutions with {de Sitter} cores and
  {Hagedorn} fluid.
\newblock {\em Class. Quantum Grav.}, 42:025024, 2025.

\bibitem{KonoplyaZhidenko2026}
Roman~A. Konoplya and Alexander Zhidenko.
\newblock Dark matter halo as a source of regular black-hole geometries.
\newblock {\em Phys. Rev. D}, 113:043011, 2026.

\bibitem{heidari2026signatures}
N~Heidari, AA~Ara{\'u}jo~Filho, V~Vertogradov, and A~{\"O}vg{\"u}n.
\newblock Signatures of a de sitter-core black hole in ringing, transmission
  and optical appearance.
\newblock {\em Physics Letters B}, 879:140606, 2026.

\bibitem{calza2025primordial}
Marco Calz{\`a}, Davide Pedrotti, and Sunny Vagnozzi.
\newblock Primordial regular black holes as all the dark matter. i.
  time-radial-symmetric metrics.
\newblock {\em Physical Review D}, 111(2):024009, 2025.

\bibitem{bambi2023regular}
Cosimo Bambi.
\newblock {\em Regular black holes}.
\newblock Springer, 2023.

\bibitem{parvez2026exact}
Tausif Parvez and S~Shankaranarayanan.
\newblock Exact, nonsingular black holes from a phantom dbi field as primordial
  dark matter.
\newblock {\em Physical Review D}, 113(10):L101503, 2026.

\bibitem{hawking1975particle}
Stephen~W. Hawking.
\newblock Particle creation by black holes.
\newblock {\em Commun. Math. Phys.}, 43(3):199--220, 1975.
\newblock Erratum: Commun. Math. Phys. 46, 206 (1976).

\bibitem{birrell1982quantum}
N.~D. Birrell and P.~C.~W. Davies.
\newblock {\em Quantum Fields in Curved Space}.
\newblock Cambridge Monographs on Mathematical Physics. Cambridge University
  Press, Cambridge, 1982.

\bibitem{wald1994quantum}
Robert~M. Wald.
\newblock {\em Quantum Field Theory in Curved Spacetime and Black Hole
  Thermodynamics}.
\newblock University of Chicago Press, Chicago, 1994.

\bibitem{AraujoFilho:2025rwr}
A.~A. Ara{\'u}jo~Filho.
\newblock {Particle production induced by a Lorentzian non-commutative
  spacetime}.
\newblock {\em Annals Phys.}, 481:170167, 2025.

\bibitem{AraujoFilho:2025hkm}
A.~A. Ara{\'u}jo~Filho.
\newblock {How does non-metricity affect particle creation and evaporation in
  bumblebee gravity?}
\newblock {\em JCAP}, 06:026, 2025.
\newblock [Erratum: JCAP 02, E01 (2026)].

\bibitem{AraujoFilho:2024ctw}
A.~A. Ara{\'u}jo~Filho.
\newblock {Particle creation and evaporation in Kalb-Ramond gravity}.
\newblock {\em JCAP}, 04:076, 2025.

\bibitem{damour1976black}
Thibaut Damour and Remo Ruffini.
\newblock Black-hole evaporation in the {Klein--Sauter--Heisenberg--Euler}
  formalism.
\newblock {\em Phys. Rev. D}, 14(2):332--334, 1976.

\bibitem{sannan1988heuristic}
Sigurd Sannan.
\newblock Heuristic derivation of the probability distributions of particles
  emitted by a black hole.
\newblock {\em Gen. Relativ. Gravit.}, 20(3):239--246, 1988.

\bibitem{srinivasan1999particle}
K.~Srinivasan and T.~Padmanabhan.
\newblock Particle production and complex path analysis.
\newblock {\em Phys. Rev. D}, 60(2):024007, 1999.

\bibitem{parikh2000hawking}
Maulik~K. Parikh and Frank Wilczek.
\newblock Hawking radiation as tunneling.
\newblock {\em Phys. Rev. Lett.}, 85(24):5042--5045, 2000.

\bibitem{vanzo2011tunnelling}
Luciano Vanzo, Giovanni Acquaviva, and Roberto Di~Criscienzo.
\newblock Tunnelling methods and {Hawking}'s radiation: Achievements and
  prospects.
\newblock {\em Class. Quantum Grav.}, 28(18):183001, 2011.

\bibitem{kerner2008fermions}
Ryan Kerner and Robert~B. Mann.
\newblock Fermions tunnelling from black holes.
\newblock {\em Class. Quantum Grav.}, 25(9):095014, 2008.

\bibitem{kraus1995self}
Per Kraus and Frank Wilczek.
\newblock Self-interaction correction to black hole radiance.
\newblock {\em Nucl. Phys. B}, 433(2):403--420, 1995.

\bibitem{page1976emission}
Don~N. Page.
\newblock Particle emission rates from a black hole. i. massless particles from
  an uncharged, nonrotating hole.
\newblock {\em Phys. Rev. D}, 13(1):198--206, 1976.

\bibitem{ahmed2026hawking}
Faizuddin Ahmed, Ahmad Al-Badawi, and Edilberto~O Silva.
\newblock Hawking temperature, sparsity and energy emission rate of regular
  black holes supported by primordial dark matter.
\newblock {\em arXiv preprint arXiv:2605.04923}, 2026.

\bibitem{skvortsova2026massive}
Milena Skvortsova.
\newblock Massive scalar quasinormal modes of an asymptotically flat regular
  black hole supported by a phantom {Dirac--Born--Infeld} field.
\newblock {\em Ann. Phys.}, 492:170587, 2026.

\bibitem{lutfuouglu2026scalar}
Bekir~Can L{\"u}tf{\"u}o{\u{g}}lu.
\newblock Scalar, electromagnetic, and {Dirac} perturbations of regular black
  holes constituting primordial dark matter.
\newblock {\em J. Cosmol. Astropart. Phys.}, 2026(07):003, 2026.

\bibitem{abdullaev2026eikonal}
Mardon Abdullaev, Gumisbek Allambergenov, Diyorbek Rashidov, Pakhlavon Jamolov,
  and Javlon Rayimbaev.
\newblock Eikonal ringing, shadows, lensing, grey-body factors, and binding
  energy of asymptotically flat regular black holes in phantom
  {Dirac--Born--Infeld} gravity.
\newblock {\em Int. J. Gravit. Theor. Phys.}, 2(2):3, 2026.

\bibitem{fulling1989aspects}
Stephen~A Fulling.
\newblock {\em Aspects of quantum field theory in curved spacetime}.
\newblock Number~17. Cambridge University Press, 1989.

\bibitem{hollands2015quantum}
Stefan Hollands and Robert~M Wald.
\newblock Quantum fields in curved spacetime.
\newblock {\em Physics Reports}, 574:1--35, 2015.

\bibitem{parker2009quantum}
Leonard Parker and David Toms.
\newblock {\em Quantum field theory in curved spacetime: quantized fields and
  gravity}.
\newblock Cambridge University Press, 2009.

\bibitem{calmet2023quantum}
Xavier Calmet, Stephen~DH Hsu, and Marco Sebastianutti.
\newblock Quantum gravitational corrections to particle creation by black
  holes.
\newblock {\em Physics Letters B}, 841:137820, 2023.

\bibitem{parikh2004energy}
Maulik~K Parikh.
\newblock Energy conservation and hawking radiation.
\newblock {\em arXiv preprint hep-th/0402166}, 2004.

\bibitem{o69}
Ryan Kerner and Robert~B Mann.
\newblock Fermions tunnelling from black holes.
\newblock {\em Classical and Quantum Gravity}, 25(9):095014, 2008.

\bibitem{o75}
Mudassar Rehman and K~Saifullah.
\newblock Charged fermions tunneling from accelerating and rotating black
  holes.
\newblock {\em Journal of Cosmology and Astroparticle Physics}, 2011(03):001,
  2011.

\bibitem{o71}
Roberto Di~Criscienzo and Luciano Vanzo.
\newblock Fermion tunneling from dynamical horizons.
\newblock {\em Europhysics Letters}, 82(6):60001, 2008.

\bibitem{o70}
Alexandre Yale.
\newblock Exact hawking radiation of scalars, fermions, and bosons using the
  tunneling method without back-reaction.
\newblock {\em Physics Letters B}, 697(4):398--403, 2011.

\bibitem{o73}
Ryan Kerner and Robert~B Mann.
\newblock Charged fermions tunnelling from kerr--newman black holes.
\newblock {\em Physics letters B}, 665(4):277--283, 2008.

\bibitem{o74}
Alexandre Yale and Robert~B Mann.
\newblock Gravitinos tunneling from black holes.
\newblock {\em Physics Letters B}, 673(2):168--172, 2009.

\bibitem{o72}
Hui-Ling Li, Shu-Zheng Yang, Teng-Jiao Zhou, and Rong Lin.
\newblock Fermion tunneling from a vaidya black hole.
\newblock {\em Europhysics Letters}, 84(2):20003, 2008.

\bibitem{page1976particle}
Don~N Page.
\newblock Particle emission rates from a black hole. ii. massless particles
  from a rotating hole.
\newblock {\em Physical Review D}, 14(12):3260, 1976.

\bibitem{hiscock1990evolution}
William~A Hiscock and Lance~D Weems.
\newblock Evolution of charged evaporating black holes.
\newblock {\em Physical Review D}, 41(4):1142, 1990.

\bibitem{macedo2015scattering}
Caio~FB Macedo, Ednilton~S de~Oliveira, and Lu{\'\i}s~CB Crispino.
\newblock Scattering by regular black holes: planar massless scalar waves
  impinging upon a bardeen black hole.
\newblock {\em Physical Review D}, 92(2):024012, 2015.

\bibitem{macedo2016absorption}
Caio~FB Macedo, Luiz~CS Leite, and Lu{\'\i}s~CB Crispino.
\newblock Absorption by dirty black holes: Null geodesics and scalar waves.
\newblock {\em Physical Review D}, 93(2):024027, 2016.

\bibitem{anacleto2023absorption}
MA~Anacleto, FA~Brito, JAV Campos, and E~Passos.
\newblock Absorption, scattering and shadow by a noncommutative black hole with
  global monopole.
\newblock {\em The European Physical Journal C}, 83(4):298, 2023.

\bibitem{macedo2013absorption}
Caio~FB Macedo, Luiz~CS Leite, Ednilton~S Oliveira, Sam~R Dolan, and Luis~CB
  Crispino.
\newblock Absorption of planar massless scalar waves by kerr black holes.
\newblock {\em Physical Review D}, 88(6):064033, 2013.

\bibitem{dolan2009scattering}
Sam~R Dolan, Ednilton~S Oliveira, and Lu{\'\i}s~CB Crispino.
\newblock Scattering of sound waves by a canonical acoustic hole.
\newblock {\em Physical Review D—Particles, Fields, Gravitation, and
  Cosmology}, 79(6):064014, 2009.

\bibitem{futterman1986scattering}
J~AH Futterman, FA~Handler, and Richard~Alfred Matzner.
\newblock Scattering from black holes.
\newblock 1986.

\bibitem{Marco1}
Marco A.~A. Paula, Luiz C.~S. Leite, and Lu{\'\i}s C.~B. Crispino.
\newblock {Electrically charged black holes in linear and nonlinear
  electrodynamics: Geodesic analysis and scalar absorption}.
\newblock {\em Phys. Rev. D}, 102(10):104033, 2020.

\bibitem{Marco2}
Marco A.~A. de~Paula, Luiz C.~S. Leite, and Lu{\'\i}s C.~B. Crispino.
\newblock {Scattering properties of charged black holes in nonlinear and
  Maxwell{\textquoteright}s electrodynamics}.
\newblock {\em Eur. Phys. J. Plus}, 137(7):785, 2022.

\bibitem{Marco3}
Marco A.~A. de~Paula, Luiz C.~S. Leite, and Lu{\'\i}s C.~B. Crispino.
\newblock {Geodesic analysis, absorption and scattering in the static Hayward
  spacetime}.
\newblock 11 2023.

\bibitem{heidari2024scattering}
N~Heidari, Caio~FB Macedo, and AA~Ara{\'u}jo Filho.
\newblock Scattering effects of bumblebee gravity in metric-affine formalism.
\newblock {\em The European Physical Journal C}, 84(11):1221, 2024.

\bibitem{Das1996we}
Sumit~R. Das, Gary~W. Gibbons, and Samir~D. Mathur.
\newblock {Universality of low-energy absorption cross-sections for black
  holes}.
\newblock {\em Phys. Rev. Lett.}, 78:417--419, 1997.

\bibitem{heidari2025absorption}
A~{\"O}vg{\"u}n.
\newblock Absorption, scattering, geodesics, shadows and lensing phenomena of
  black holes in effective quantum gravity.
\newblock {\em Physics of the Dark Universe}, 47:101815, 2025.

\bibitem{Higuchi:2001si}
Atsushi Higuchi.
\newblock {Low frequency scalar absorption cross-sections for stationary black
  holes}.
\newblock {\em Class. Quant. Grav.}, 18:L139, 2001.
\newblock [Addendum: Class.Quant.Grav. 19, 599 (2002)].

\bibitem{leite2017scalar}
Luiz~CS Leite, Carolina~L Benone, and Lu{\'\i}s~CB Crispino.
\newblock Scalar absorption by charged rotating black holes.
\newblock {\em Physical Review D}, 96(4):044043, 2017.

\bibitem{baptista2025scattering}
Arturo~Quispe Baptista and ML~Pe{\~n}afiel.
\newblock Scattering and absorption of massless scalar waves by a modmax black
  hole.
\newblock {\em Physical Review D}, 112(2):024037, 2025.

\bibitem{heidari2026particle}
N~Heidari, PHM Barros, et~al.
\newblock Particle production, absorption, scattering, and geodesics in a
  schwarzschild--hernquist black hole.
\newblock {\em The European Physical Journal C}, 86(5):486, 2026.

\bibitem{dolan2013scattering}
Sam~R Dolan and Ednilton~S Oliveira.
\newblock Scattering by a draining bathtub vortex.
\newblock {\em Physical Review D—Particles, Fields, Gravitation, and
  Cosmology}, 87(12):124038, 2013.

\bibitem{anacleto2020absorption}
MA~Anacleto, FA~Brito, JAV Campos, and E~Passos.
\newblock Absorption and scattering by a self-dual black hole.
\newblock {\em General Relativity and Gravitation}, 52(10):100, 2020.

\bibitem{leite2019black}
Luiz~CS Leite, Caio~FB Macedo, and Lu{\'\i}s~CB Crispino.
\newblock Black holes with surrounding matter and rainbow scattering.
\newblock {\em Physical Review D}, 99(6):064020, 2019.

\bibitem{sanchez1978elastic}
Norma S{\'a}nchez.
\newblock Elastic scattering of waves by a black hole.
\newblock {\em Physical Review D}, 18(6):1798, 1978.

\bibitem{yennie1954phase}
DR~Yennie, D\_~G\_ Ravenhall, and RN~Wilson.
\newblock Phase-shift calculation of high-energy electron scattering.
\newblock {\em Physical Review}, 95(2):500, 1954.

\bibitem{dolan2006fermion}
Sam Dolan, Chris Doran, and Anthony Lasenby.
\newblock Fermion scattering by a schwarzschild black hole.
\newblock {\em Physical Review. D, Particles Fields}, 74(6), 2006.

\bibitem{AraujoFilho:2025hnf}
A.~A. Ara{\'u}jo~Filho, N.~Heidari, I.~P. Lobo, and V.~B. Bezerra.
\newblock {Gravitational signatures of a nonlinear electrodynamics in f(R,T)
  gravity}.
\newblock {\em JCAP}, 09:015, 2025.
\newblock [Erratum: JCAP 01, E01 (2026)].

\bibitem{AraujoFilho:2024mvz}
A.~A. Ara{\'u}jo~Filho, N.~Heidari, and Ali {\"O}vg{\"u}n.
\newblock {Geodesics, accretion disk, gravitational lensing, time delay, and
  effects on neutrinos induced by a non-commutative black hole}.
\newblock {\em JCAP}, 06:062, 2025.

\bibitem{Filho:2024tgy}
A.~A.~Ara{\'u}jo Filho.
\newblock {Antisymmetric tensor influence on charged black hole lensing
  phenomena and time delay}.
\newblock {\em JHEAp}, 47:100401, 2025.

\bibitem{Filho:2024isd}
A.~A.~Ara{\'u}jo Filho, J.~R. Nascimento, A.~Yu. Petrov, and P.~J.
  Porf{\'\i}rio.
\newblock {Gravitational lensing by a Lorentz-violating black hole}.
\newblock {\em Eur. Phys. J. Plus}, 140(11):1117, 2025.

\bibitem{Filho:2021zuf}
A.~A.~Ara{\'u}jo Filho and A.~Yu. Petrov.
\newblock {Bouncing universe in a heat bath}.
\newblock {\em Int. J. Mod. Phys. A}, 36(34n35):2150242, 2021.

\bibitem{furtado2023thermal}
J~Furtado, H~Hassanabadi, JAAS Reis, et~al.
\newblock Thermal analysis of photon-like particles in rainbow gravity.
\newblock {\em arXiv preprint arXiv:2305.08587}, 2023.

\bibitem{AraujoFilho:2026tvy}
A.~A. Ara{\'u}jo~Filho.
\newblock {Thermodynamic and statistical properties of a multifractional
  modified dispersion relation via the grand-canonical ensemble}.
\newblock 5 2026.

\end{thebibliography}
	\bibliographystyle{unsrt}
	
\end{document}